\documentclass[journal,twoside,letterpaper,web]{ieeecolor}
\usepackage{eso-pic}
\usepackage{generic}
\usepackage{cite}
\usepackage{amsmath,amssymb,amsfonts}
\usepackage{algorithmic}
\usepackage{graphicx}
\usepackage{algorithm,algorithmic}
\usepackage{textcomp}
\usepackage{bm}
\usepackage[caption=false]{subfig}
\usepackage{multirow}
\usepackage{tikz}
\usetikzlibrary{patterns, calc, arrows.meta, positioning, decorations.pathmorphing}
\usepackage{calligra}
\usepackage{xcolor}
\usepackage{hyperref}
\hypersetup{hidelinks=true,colorlinks=false, linkcolor=black, filecolor=magenta, urlcolor=cyan}

\newtheorem{assumption}{Assumption}
\newtheorem{remark}{Remark}
\newtheorem{note}{Note}
\newtheorem{property}{Property}

\def\BibTeX{{\rm B\kern-.05em{\sc i\kern-.025em b}\kern-.08em
    T\kern-.1667em\lower.7ex\hbox{E}\kern-.125emX}}
\begin{document}
	
\AddToShipoutPictureBG*{%
	\put(0,020){
		\hspace*{\dimexpr0.043\paperwidth\relax}
		\parbox{.94\paperwidth}{\footnotesize This work has been submitted to the IEEE for possible publication. Copyright may be transferred without notice, after which this version may no longer be accessible.}%
}}
\title{Stochastic Model Predictive Control with Online Risk Allocation and Feedback Gain Selection}
\author{Filipe Marques Barbosa and Johan L{\"o}fberg
\thanks{This work was supported by VINNOVA Competence Center Link-SIC.}
\thanks{Filipe Marques Barbosa and Johan L{\"o}fberg are with the Division of Automatic Control, Department of Electrical Engineering, Link{\"o}ping University, Link{\"o}ping, Sweden (filipe.barbosa@liu.se, johan.lofberg@liu.se)}}

\maketitle

\begin{abstract}
Stochastic Model Predictive Control addresses uncertainties by incorporating chance constraints that provide probabilistic guarantees of constraint satisfaction. However, simultaneously optimizing over the risk allocation and the feedback policies leads to intractable nonconvex problems. This is due to (i) products of functions involving the feedback law and risk allocation in the deterministic counterpart of the chance constraints, and (ii) the presence of the nonconvex Gaussian quantile (probit) function. Existing methods rely on two-stage optimization, which is nonconvex. To address this, we derive disjunctive convex chance constraints and select the feedback law from a set of precomputed candidates. The inherited compositions of the probit function are replaced with power- and exponential-cone representable approximations. The main advantage is that the problem can be formulated as a mixed-integer conic optimization problem and efficiently solved with off-the-shelf software. Moreover, the proposed formulations apply to general chance constraints with products of exclusive disjunctive and Gaussian variables. The proposed approaches are validated with a path-planning application.
\end{abstract}


\section{Introduction}
\IEEEPARstart{M}{odel} predictive control (MPC) is an advanced technique to control multivariable dynamic systems under constraints with applications in various engineering fields. At each sampling instant, an optimal control problem (OCP) is solved over a finite horizon, giving a sequence of \textit{control inputs}. The first input of this sequence is applied to the system, and the process repeats in a receding horizon fashion.  This provides an ``implicit'' feedback action to handle system uncertainties and disturbances.

Although the receding-horizon implementation offers a degree of robustness in classical MPC, its deterministic formulation makes it fundamentally inadequate for systematically addressing uncertainties. With this in mind, robust MPC (RMPC) approaches consider uncertainties using deterministic, bounded, set-membership descriptions. Early works on RMPC employed min-max formulations to compute control inputs that guarantee constraint satisfaction under all admissible disturbances \cite{Bemporad}. Min-max approaches are, however, overly conservative and may lead to infeasibility. To mitigate these limitations, affine disturbance feedback (discussed later) and tube-based RMPC approaches were introduced \cite{Lofberg2003, Goulart2006,Langson2004}, where constraints are tightened to ensure that the system remains within a ``tube'' around the nominal trajectory. Nonetheless, RMPC formulations can still be overly conservative, as they consider worst-case bounds, often neglecting the statistical properties of disturbances.

A natural extension of RMPC with stochastic descriptions of uncertainties is to incorporate the probability of disturbance occurrences explicitly. This leads to stochastic MPC (SMPC), which exploits the statistical characterization of uncertainties by replacing hard worst-case constraints with soft \textit{chance constraints} that must be satisfied with at least a specified probability level. This enables a systematic trade-off between robustness and performance, resulting in less conservative solutions \cite{Farina2016}.

Ensuring that the chance constraints are satisfied simultaneously over the entire prediction horizon results in formulations with \textit{joint chance constraints}. Although more intuitive for formulating SMPC problems, joint chance constraints are generally nonconvex and computationally intractable \cite{Bertsimas2011}. To address this, convex approximations are used to obtain tractable surrogates. These can be derived using sampling-based methods \cite{Campi2009,Blackmore2010} or through analytical safe approximations \cite{Bertsimas2005,Calafiore2006,Blackmore2006,Nemirovski2007}. Sampling-based approaches only provide probabilistic guarantees of constraint satisfaction and are computationally demanding due to the large number of samples required for accuracy. On the other hand, analytical safe approximations derive the chance constraint deterministically, yielding convex surrogates whose feasible set is contained within that of the original problem, ensuring that any feasible solution to the approximate problem is also feasible for the original joint chance constraints.

A standard method for obtaining an analytical safe approximation of a joint chance constraint is to decompose it into \textit{individual chance constraints} and bound their probabilities of violation using Boole's inequality, which yields tighter approximations \cite{Blackmore2009}. The main challenge lies in determining how the total allowable risk of constraint violation is distributed along the prediction horizon, i.e., the \textit{risk allocation}. The risk allocation can either be fixed a priori, e.g., \cite{Li2022,Blackmore2011,Zhang2021,Knaup2023,Fink2024}, or treated as a decision variable. Though fixing the risk allocation simplifies the optimization, it often leads to conservative solutions, as the optimal risk allocation may vary as the system's dynamics evolve. Conversely, treating the individual risks as decision variables can reduce this conservatism.

A problem with treating the risk allocation as decision variables is the handling of the quantile (inverse cumulative distribution) function, which is obtained when deriving a deterministic expression for the probabilistic constraints. Some works addressed this by using the Cantelli-Chebyshev inequality, assuming that only the disturbance's first and second moments are known \cite{Paulson2017,Li2022,Wang2022}. Nonetheless, assuming Gaussian disturbances is desired in many applications. In such cases, although tractable, the Cantelli-Chebyshev inequality introduces significant conservativeness. See \cite{Heirung2018} for a discussion. On the other hand, directly using the quantile function of the standard Gaussian distribution (probit function) in the OCP formulation yields a general nonlinear, nonconvex optimization problem involving a non-elementary function, making it computationally expensive or intractable.

With this in mind, Barbosa and L{\"o}fberg \cite{Barbosa2025} proposed replacing the probit function with an exponential-cone representable approximation to achieve tractability with reduced conservativeness. However, only stable open-loop SMPC was considered, and, for unstable models, increasingly conservative control inputs are produced due to the disturbance propagation throughout the prediction horizon.

Beyond the handling of the risk allocation, when accounting for disturbances in constrained OCPs, a better solution is to optimize over the admissible set of feedback policies rather than the open-loop input sequences. One option is to precompute stabilizing linear feedback laws offline. Unfortunately, it is not always obvious how to best choose the control laws to minimize conservativeness, as conditions and performance may vary as the system evolves. An alternative is to optimize over the affine state feedback policies online, but the resulting set of admissible decision variables is nonconvex in general \cite{Goulart2006}. To circumvent this, an affine \textit{disturbance feedback parameterization} of the control policies was proposed in \cite{Lofberg2003}. This parameterization is guaranteed to be convex and is equivalent to the state feedback parameterization \cite{Goulart2006}. Consequently, it has since been widely adopted in constrained optimal control problems under disturbances. Examples of application in the context of SMPC include \cite{Oldewurtel2008,Hokayem2012,Paulson2017,Zhang2021}.

Yet, simultaneously optimizing over both the feedback law and risk allocation leads to nonconvex problems. This is caused by
the product of their respective functions in the deterministic counterpart of the chance constraints. A common approach to address this problem is to solve a two-stage (bilevel) optimization, where the upper-stage optimizes the risk allocation and the lower-stage optimizes over the feedback laws \cite{MasahiroOno2008,Vitus2011,Paulson2017,Pilipovsky2021}. However, this approach offers no guarantee of convergence, even to a local optimum. Each stage solves its respective problem with the other held fixed.

We argue that a better approach is to precompute a sufficiently rich set of disturbance feedback laws and optimize over their selection. This selection is modeled using binary indicator variables, resulting in a mixed-integer optimization (MIO) problem solved via the branch-and-bound (B\&B) algorithm. B\&B is arguably the most important framework for solving MIO problems efficiently. It systematically explores the solution space by successively solving a series of \textit{continuous relaxations} of the original problem, creating branches to refined relaxations and pruning suboptimal regions. 

However, simply recasting the SMPC problem as a mixed-integer optimization problem does not resolve the core issue here. The deterministic counterparts of the chance constraints remain nonconvex due to products of functions now involving the feedback selection and risk allocation -- i.e., disjunctive variables multiplied by Gaussian random variables. Thus, they must be reformulated as convex constraints. Yet, any attempt of reformulation will inevitably inherit a nonconvex composition involving the probit function. This issue can be addressed by replacing the nonconvex function composition with a nonsymmetric conic approximation, following the approach proposed in \cite{Barbosa2025}.

Nonsymmetric cones broaden the general framework of conic optimization by allowing a wider class of convex problems to be formulated within the conic structure \cite{Chares2009}. In terms of practical importance, the three-dimensional power- and exponential-cones are perhaps the most important nonsymmetric cones. They are convex by construction and retain key interior-point properties required for tractable conic optimization. Thus, specialized solvers can exploit their structure to achieve good numerical performance. Early implementations demonstrating this include ECOS \cite{Serrano2015} (only for exponential cones) and SCS \cite{ODonoghue2016}. Following this, Dahl and Andersen \cite{Dahl2021} generalized the algorithm proposed by Nesterov and Todd \cite{Nesterov1997,Nesterov1998} to handle the nonsymmetric power and exponential cones by incorporating the primal-dual scalings proposed by Tun\c{c}el \cite{Tunel2001}.
This results in a practical implementation capable of handling large-scale problems efficiently. This implementation is available in MOSEK ApS\footnote{Available at \url{https://www.mosek.com/}} and is used in this work. 


In this paper, we propose three disjunctive convex formulations of chance constraints where both the disturbance feedback law and risk allocation are optimized over. These formulations are convex within low-risk regions but inherently contain a non-elementary, nonconvex composition of the probit function. Using these compositions directly in the optimization problem would still lead to general nonlinear, nonconvex, and intractable continuous relaxations. To address this, we replace these compositions with power- and exponential-cone representable approximations, following the approach in \cite{Barbosa2025}, which reduce conservativeness and yield tractable convex relaxations. The main advantage is that the resulting OCP can be expressed as a mixed-integer conic optimization problem, for which efficient solvers and algorithmic frameworks are available. Furthermore, the proposed approach can be generalized to chance constraints involving products of mutually exclusive binary variables and Gaussian random variables, where only one combination is selected.


\paragraph*{Notation} 
For a vector \(\bm x \in \mathbb{R}^{n}\), the weighted norm is denoted by \(\lVert \bm x \rVert_{\bm Q} = \sqrt{\bm x^{\top} \bm Q \bm x}\), where \(\bm Q \succeq 0\). The \(i\)-th vector or matrix in a sequence is denoted as \(\bm x_i\) or \(\bm A_i\), whereas the \(i\)-th row of a matrix \(\bm F\) is denoted as \(\bm F_{(i)}\). Superscripts may indicate exponents, dimensions, or specific characteristics, with the intended meaning clear from context. Scalar variables in the Latin alphabet (e.g., \(t\), \(y\), and \(z\)) are not reserved and should be interpreted according to context.

\section{Preliminaries}
\label{sec:preliminaries}

We begin by introducing the affine disturbance feedback parameterization, a crucial concept for online optimization of feedback policies in stochastic model predictive control.

\subsection{Stochastic Linear System}

Consider the class of linear discrete-time systems with additive disturbances
\begin{equation}
	\bm{x}_{i+1} =  \bm{A} \bm{x}_{i} + \bm{B} \bm{u}_{i} + \bm{G} \bm{\omega}_{i}
	\label{eq:state-space1}
\end{equation}
with the state vector \(\bm{x}_{i} \in \mathbb{R}^{n_{x}}\), the control input \(\bm{u}_{i} \in \mathbb{R}^{n_{u}}\), and the stochastic disturbance \(\bm{\omega}_{i} \in \mathbb{R}^{n_{\omega}}\). For notational  convenience, the prediction of the system dynamics over a finite horizon \(N \in \mathbb{N}\) is represented as
\begin{equation}
	\mathbf{X} = \bm{\mathcal{A}} \bm{x}_{0} + \bm{\mathcal{B}} \mathbf{U} + \bm{\mathcal{G}} \mathbf{W}
	\label{eq:state-space-open}
\end{equation}
with stacked vectors defining the sequence of state predictions, control inputs, and stochastic disturbances, respectively, as 
\begin{align*}
	\mathbf{X} &= [\bm{x}_{0}^\top, \bm{x}_{1}^\top, \dots, \bm{x}_{N}^\top]^\top \in \mathbb{R}^{(N+1)n_{x}}\\
	\mathbf{U} &= [\bm{u}_{0}^\top,\bm{u}_{1}^\top, \dots, \bm{u}_{N-1}^\top]^\top \in \mathbb{R}^{Nn_{u}}\\
	\mathbf{W} &= [\bm{\omega}_{0}^\top, \bm{\omega}_{1}^\top, \dots, \bm{\omega}_{N-1}^\top]^\top \in \mathbb{R}^{Nn_{\omega}}
\end{align*}
and the matrices \(\bm{\mathcal{A}} \in \mathbb{R}^{(N+1)n_{x} \times n_{x}}\), \(\bm{\mathcal{B}} \in \mathbb{R}^{(N+1)n_{x} \times Nn_{u}}\), and \(\bm{\mathcal{G}} \in \mathbb{R}^{(N+1)n_{x} \times Nn_{\omega}}\) defined as
\begin{gather}
	\bm{\mathcal{A}} = 
	\begin{bmatrix}
		\bm{I}\\
		\bm{A}\\
		\bm{A}^{2}\\
		\vdots\\
		\bm{A}^{N}
	\end{bmatrix},
	\bm{\mathcal{B}} = 
	\begin{bmatrix}
		\bm{0} & \bm{0} & \dots & \bm{0}\\
		\bm{B} & \bm{0} & \dots & \bm{0}\\
		\bm{A}\bm{B} & \bm{B} & \dots & \bm{0}\\
		\vdots & \vdots & \ddots & \vdots\\
		\bm{A}^{N-1}\bm{B} & \bm{A}^{N-2}\bm{B} & \dots & \bm{B}
	\end{bmatrix}, \nonumber \\
	\bm{\mathcal{G}} = 
	\begin{bmatrix}
		\bm{0} & \bm{0} & \dots & \bm{0}\\
		\bm{G} & \bm{0} & \dots & \bm{0}\\
		\bm{A}\bm{G} & \bm{G} & \dots & \bm{0}\\
		\vdots & \vdots & \ddots & \vdots\\
		\bm{A}^{N-1}\bm{G} & \bm{A}^{N-2}\bm{G} & \dots & \bm{G}
	\end{bmatrix}.
	\label{eq:linear-map-matrix}
\end{gather}

\begin{assumption}
	The disturbances are assumed to be independent and identically distributed (i.i.d.) standard Gaussian random variables, i.e., \(\bm{\omega} \sim \mathcal{N}(\bm 0, \bm I)\).
\end{assumption}
\begin{note}
	We consider standard Gaussian disturbances to simplify the notation. This assumption is not restrictive, since the methods described in this paper apply equally to Gaussian disturbances with nonzero mean and non-unit variance, as long as their statistics are known.
\end{note}

When accounting for disturbances in constrained optimal control problems, open-loop input sequences may lead to excessive conservativeness as well as infeasibility or instability. Therefore, a better alternative is to optimize over admissible state feedback predictions.

\subsection{Disturbance Feedback Parameterization}
\label{sec:disturbance-feedback}

Feedback predictions can be employed in system \eqref{eq:state-space1} by parameterizing the future control sequence in terms of the future states as
\begin{equation}
	\bm u_{i} = \sum_{j=0}^{i}\bm L_{i,j} \bm x_{j} + \bm v_{i}, ~ \forall i = 1,\dots,N-1,
	\label{eq:control-input-state}
\end{equation}
where \(\bm{L}_{i,j} \in \mathbb{R}^{n_{u} \times n_{x}}\) is the feedback gain matrix and \(\bm{v}_{i} \in \mathbb{R}^{n_{u}}\) is the nominal control input (also known as the feedforward term). For notational convenience, we also define the vector
\begin{equation}
	\mathbf{V} = [\bm{v}_{0}^\top, \bm{v}_{1}^\top, \dots, \bm{v}_{N-1}^\top]^\top \in \mathbb{R}^{Nn_{u}}
\end{equation}
and the lower triangular block matrix \(\bm{\mathcal{L}} \in \mathbb{R}^{Nn_{u} \times (N+1)n_{x}}\) as
\begin{equation}
	\bm{\mathcal{L}} = 
	\begin{bmatrix}
		\bm{L}_{0,0} & \bm{0} & \dots & \bm{0} \\
		\vdots & \ddots & \ddots & \vdots \\
		\bm{0} & \dots & \bm{L}_{N-1,N-1} & \bm{0}
	\end{bmatrix},
	\label{eq:state-feedback-block}
\end{equation}
and \eqref{eq:control-input-state} can be written in stacked form as \(\mathbf{U} = \bm{\mathcal{L}}\mathbf{X} + \mathbf{V}\).

A common approach is to fix a stabilizing \(\bm{\mathcal{L}}\) and regard \(\ \mathbf{V}\) as decision variables. However, it is not always obvious how to choose the feedback gain offline so as to minimize conservativeness. Therefore, a natural approach is to also optimize over the selection of \(\bm{\mathcal{L}}\) and have more degrees of freedom.

However, this generally leads to an intractable, nonconvex set of admissible control parameters \cite{Goulart2006}. To circumvent this, we adopt the affine disturbance feedback policy introduced by L{\"{o}}fberg \cite{Lofberg2003} and parameterize the control sequence directly in terms of the disturbance
\begin{equation}
	\bm{u}_{i} = \sum_{j=0}^{i-1} \bm{M}_{i,j} \bm{\omega}_{j} + \bm{v}_{i}, ~ \forall i = 1,\dots,N-1,
	\label{eq:control-input}
\end{equation}
where \(\bm{M}_{i,j} \in \mathbb{R}^{n_{u} \times n_{\omega}}\) describes how \(\bm{u}_{i}\) uses \(\bm{\omega}_{j}\). Note that this parameterization is causal, i.e., \(\bm u_i\) is affected only by \(\bm \omega_{j}\), \(j < i\). We then define the strictly lower triangular block matrix \(\bm{\mathcal{M}} \in \mathbb{R}^{Nn_{u} \times Nn_{\omega}}\) as
\begin{equation}
	\bm{\mathcal{M}} = 
	\begin{bmatrix}
		\bm{0} & \bm{0} & \dots & \bm{0} \\
		\bm{M}_{1,0} & \bm{0} & \dots & \bm{0} \\
		\vdots & \ddots & \ddots & \vdots \\
		\bm{M}_{N-1,0} & \bm{M}_{N-1,N-2} & \dots & \bm{0}
	\end{bmatrix}.
\end{equation}
This yields the control inputs in stacked form as 
\begin{equation}
	\mathbf{U} = \bm{\mathcal{M}}\mathbf{W} + \mathbf{V}\\
	\label{eq:input-disturbance-feedback}
\end{equation}
and \eqref{eq:state-space-open} becomes
\begin{equation}
	\mathbf{X} = \bm{\mathcal{A}} \bm{x}_{0} + \bm{\mathcal{B}} \mathbf{V} + \left( \bm{\mathcal{G}} + \bm{\mathcal{B}}\bm{\mathcal{M}} \right) \mathbf{W}.
	\label{eq:state-space-closed}	
\end{equation}

The main advantage of this parameterization is that it allows the formulation of convex robust MPC problems with online optimization of the feedback predictions. Moreover, Goulart \textit{et al.} \cite{Goulart2006} showed that, for every admissible disturbance feedback policy, there exists an admissible state feedback policy that yields the same state and control input sequences for all disturbance realizations. This policy can be constructed with
\begin{equation}
	\bm{\mathcal{M}} = \bm{\mathcal{L}}\left(\bm I -\bm{\mathcal{B}}\bm{\mathcal{L}} \right)^{-1}\bm{\mathcal{G}}.
	\label{eq:disturbance-feedback-equivalence}
\end{equation}

Robust formulations can, however, be overly conservative, as they account for the worst-case disturbance and often disregard the statistical properties of the uncertainty. Thus, a less conservative alternative is to account for the disturbances in a probabilistic sense, leading to stochastic formulations.

\section{Chance Constrained Optimization}
\label{sec:chance-constraints}

Instead of accounting for the worst-case uncertainty and guaranteeing constraint satisfaction under all possible disturbances, we use the stochastic descriptions of the uncertainties and ensure that the constraints containing stochastic parameters remain within probabilistic bounds. This way, the control inputs are computed to guarantee the constraint satisfaction with at least a predefined probability level. Thus, the probability \(\mathbb{P}\) of satisfying the constraints over the entire prediction horizon is defined as a \textit{joint chance constraint}
\begin{equation}
	\mathbb{P}\left( \mathbf{X} \in \mathcal{X} \right) \geq 1 - \xi,
	\label{eq:joint-cc}
\end{equation}
where \(\mathcal{X}\) is the feasible region and \(\xi\) is the joint risk of state constraint violation. This leads to SMPC problems formulated as 
\begin{equation}
	\begin{aligned}
		& \underset{\mathbf{V}}{\mathrm{minimize}}
		&& \mathbb{E}\,[\mathrm{J}(\mathbf{X},\mathbf{U},\mathbf{W})]\\
		& \mathrm{subject ~ to}
		&& \mathbf{X} = \bm{\mathcal{A}} \bm{x}_{0} + \bm{\mathcal{B}} \mathbf{V} + \left( \bm{\mathcal{G}} + \bm{\mathcal{B}}\bm{\mathcal{M}} \right) \mathbf{W}\\
		&&& \mathbb{P}\left(\mathbf{X} \in \mathcal{X} \right)\geq 1-\xi\\
		&&& \mathbb{P}\left( \mathbf{U} \in \mathcal{U} \right) \geq 1-\zeta
	\end{aligned}
	\label{eq:SMPC-general}
\end{equation}
where the expectation of the cost \(\mathbb{E}\,[\mathrm{J}(\mathbf{X},\mathbf{U},\mathbf{W})]\) and the region \(\mathcal{U}\) are assumed convex, and \(\zeta\) is the risk of input constraint violation.

\begin{note}
	As a consequence of using the disturbance feedback parameterization, the control inputs now contain stochastic components. Therefore, imposing hard constraints on \(\bm u_{i}\) is not possible; because the underlying Gaussian distribution has unbounded support, any value can be obtained with vanishing probability of violation.
\end{note}

Evaluating joint chance constraints, in principle, requires the computation of a multivariate integral, which escalates in difficulty for higher dimensions. As a result, these constraints are generally nonconvex, and an exact, tractable representation may not exist. To circumvent this, we use Boole's inequality to decompose the joint chance constraints into individual constraints and bound the probability of violation.

\subsection{Boole's Inequality}
\label{sec:boole-inequality}

Boole's inequality states that the probability of at least one event occurring is less than or equal to the sum of the probabilities of the individual events, that is,
\begin{equation}
	\mathbb{P}\left(\bigwedge_{\ell=1}^{N_{C}} \mathbf{X}_{(\ell)} \not\in \mathcal{X}\right) \leq \sum_{\ell=1}^{N_{C}}\mathbb{P}\left(\mathbf{X}_{(\ell)}\not\in \mathcal{X}\right)
	\label{eq:boole_inequality}
\end{equation}
with \(N_{c}\in\mathbb{N}\). This means that if an individual probability of constraint violation is bounded by \(\mathbb{P}(\mathbf{X}_{(\ell)} \not\in \mathcal{X})\leq \gamma_\ell\), then the probability of at least one constraint violation is bounded by \(\sum_{\ell=1}^{N_{C}}\gamma_\ell\). Hence, the joint constraints in (\ref{eq:joint-cc}) can be replaced by \(N_{C}\) individual chance constraints as
\begin{equation}
	\mathbb{P}\left(\mathbf{X}_{(\ell)}\in\mathcal{X}\right)\geq 1-\gamma_\ell,~\ell=1,\dots, N_{C}
	\label{eq:individial-cc}
\end{equation}
and satisfy (\ref{eq:boole_inequality}) when the \textit{risk allocation} \(\gamma_\ell \in [0,\xi]\) is bounded by
\begin{equation}
	\sum_{\ell=1}^{N_{C}}\gamma_\ell \leq \xi.
	\label{eq:risk-bound}
\end{equation}

Following this, the individual probabilistic constraints can be represented analytically and used in an optimization framework.

\begin{note}
	All probabilistic constraints in the remainder of this paper are individual chance constraints, and hence \eqref{eq:risk-bound} applies. We omit it for the sake of simplicity and keeping the focus on the presented approach.
\end{note}

\subsection{Deterministic Representation}

Individual linear chance constraints with additive Gaussian disturbances can be derived into \textit{exact} deterministic constraints. To do it, we express the chance constraints in (\ref{eq:individial-cc}) explicitly as 
\begin{equation}
	\mathbb{P}\left( \bm H_{(\ell)} \left(\bm{\mathcal{A}} \bm x_{0} \! + \! \bm{\mathcal{B}} \! \mathbf{V} \! + \! \left( \bm{\mathcal{G}} \! + \! \bm{\mathcal{B}} \bm{\mathcal{M}} \right)\! \mathbf{W}\right) \leq h_{\ell}\right) \geq 1 - \gamma_{\ell},
	\label{eq:chance-explict}
\end{equation}
where \(\bm{H} \in \mathbb{R}^{n_{h}\times(N+1)n_{x}}\) is a constraint matrix and \(h \in \mathbb{R}\) is constant. To simplify notation, we write \eqref{eq:chance-explict} alternatively as
\begin{equation}
	\mathbb{P}( f_{\ell}(\mathbf{V}) + \bm c_{\ell}^{\top}\left(\bm{\mathcal{M}}\right) \mathbf{W} \leq 0)\geq 1-\gamma_{\ell}
	\label{eq:chance}
\end{equation}
where 
\begin{align}
	f_{\ell}(\mathbf{V}) &= \bm H_{(\ell)} \left(\bm{\mathcal{A}} \bm x_{0} + \bm{\mathcal{B}} \mathbf{V} \right)-h_{\ell} \text{ and}\\
	\bm c_{\ell}^{\top}\left(\bm{\mathcal{M}}\right) &= \bm H_{(\ell)} \left(\bm{\mathcal{G}} +  \bm{\mathcal{B}} \bm{\mathcal{M}}\right).
\end{align}

Thus, the exact deterministic representation (without approximation) of \eqref{eq:chance} becomes
\begin{equation}
	f_{\ell}(\mathbf{V}) + \lVert \bm c_{\ell}\left(\bm{\mathcal{M}}\right)  \rVert\Phi^{-1}(1-\gamma_{\ell}) \leq 0,
	\label{eq:analytical-cc-individial}
\end{equation}
where \(\Phi^{-1}\) is the probit function\footnote{The probit function is the inverse of the cumulative distribution function of the standard Gaussian distribution.}. This result is well-known in the literature, and we refer to \cite{Schwarm1999} for more details.

It should be noted that the deterministic representation obtained for the upper bound chance constraints can likewise be obtained for lower bound chance constraints, resulting in
\begin{multline}
	\mathbb{P}\left(f_{\ell}(\mathbf{V}) + \bm c_{\ell}^{\top}\left(\bm{\mathcal{M}}\right) \mathbf{W} \geq 0\right)\geq 1-\gamma_{\ell} \implies\\
	\lVert \bm c_{\ell}\left(\bm{\mathcal{M}}\right) \rVert\Phi^{-1}(1-\gamma_{\ell}) -  f_{\ell}(\mathbf{V}) \leq 0.
	\label{eq:analytical-cc-individial-upper}
\end{multline}

\subsection{Risk allocation}
\label{sec:risk-allocation}

Achieving improved control performance may require operating the system closer to its constraints, which comes at the cost of increasing the risk of constraint violations. Hence, a core aspect of SMPC with chance constraints is to determine how much each prediction \(\mathbf{X}_{(\ell)}\) must back off from \(h_{\ell}\) so that the individual chance constraints are satisfied. This is determined by the choice of values of \(\gamma_{\ell}\), i.e., how the risk is allocated.

A straightforward way of selecting the \textit{risk allocation} is to fix each \(\gamma_{\ell}\) a priori. However, though this simplifies the optimization problem, it yields conservative solutions. Consider, for instance, a vehicle moving in an environment with regions to be avoided. The risk allocation would then be the same, regardless of whether these regions are far away or too close to the vehicle. In such situations, it is desirable to trade off the risk allocation with another performance criterion. 

Thus, we treat the risk allocation as decision variables to reduce conservativeness. However, since we also want to optimize over \(\bm{\mathcal{M}}\), the product \(\lVert \bm c\left(\bm{\mathcal{M}}\right)  \rVert\Phi^{-1}(1-\gamma)\) in \eqref{eq:analytical-cc-individial} and \eqref{eq:analytical-cc-individial-upper} remains nonconvex even for \(\gamma\in [0,0.5]\) (the region where \(\Phi^{-1}(\cdot)\) is convex). As a compromise between a simple convex formulation with a fixed choice of \(\bm{\mathcal{M}}\), and the fully nonconvex case with \(\bm{\mathcal{M}}\) as a decision variable, we propose to optimize over a predefined finite set of feedback laws through mixed-integer conic formulations.

\subsection{Disjunctive Chance Constraints}
\label{sec:disjunctive-chance}

We want to optimize over the selection of a feedback law that will be used over the prediction horizon, i.e., decide on one feedback law at every sampling instant and use it over the entire prediction horizon. Thus, various feedback laws \(\bm{L}_{k}\) are computed using, for instance, linear-quadratic controllers with different tuning parameters so that some perform well in specific situations, while others are better under different conditions.

To do this, introduce the indicator variables \(\delta_k \in \{0,1\}\) representing the binary choices for the feedback laws. The selection is then made as
\begin{equation}
	\bm{L} = \sum_{k=1}^{N_{F}} \delta_k \bm{L}_k 
\end{equation}
with \(\sum \delta_k = 1\), where \(N_{F} \in \mathbb{N}\) is the number of pre-computed feedback laws. This way, we construct \(N_{F}\) state feedback block matrices, as given in \eqref{eq:state-feedback-block}, and use them to obtain the causal affine disturbance feedback given by \eqref{eq:disturbance-feedback-equivalence} as
\begin{equation}
	\bm{\mathcal{M}}(\delta) = \sum_{k=1}^{N_{F}} \delta_{k}\bm{\mathcal{L}}(\bm{L}_k)\left(\bm I -\bm{\mathcal{B}}\bm{\mathcal{L}}(\bm{L}_k) \right)^{-1}\bm{\mathcal{G}}.
	\label{eq:disturbance-feedback-equivalence-selection}
\end{equation}
In other words, we think of the feedback in terms of the disturbances, with \(\bm L_{k}\) as the objects we select from. Consequently, the control inputs are given by
\begin{equation}
	\mathbf{U} = \bm{\mathcal{M}}(\delta)\mathbf{W} + \mathbf{V}
	\label{eq:input-disturbance-feedback-selection}
\end{equation}
and the state predictions in \eqref{eq:state-space-closed} become
\begin{equation}
	\mathbf{X} = \bm{\mathcal{A}} \bm{x}_{0} + \bm{\mathcal{B}} \mathbf{V} + \left( \bm{\mathcal{G}} + \bm{\mathcal{B}}\bm{\mathcal{M}}(\delta) \right) \mathbf{W}.
	\label{eq:state-space-closed-selection}	
\end{equation}
The chance constraints now involve the stochastic expressions \eqref{eq:input-disturbance-feedback-selection} and \eqref{eq:state-space-closed-selection}, and \eqref{eq:chance} becomes a \textit{disjunctive chance constraint}
\begin{equation}
	\mathbb{P}\left( f_{\ell} (\mathbf{V}) + \bm c_{\ell}^{\top}(\delta) \mathbf{W} \leq 0\right) \geq 1-\gamma_{\ell}
	\label{eq:chance-selection}
\end{equation}
which evaluates to
\begin{equation}
	f_{\ell}(\mathbf{V}) + \lVert \bm c_{\ell} (\delta)\rVert\Phi^{-1}(1-\gamma_{\ell}) \leq 0
	\label{eq:analytical-cc-individial-disjoint}
\end{equation}
with 
\begin{equation}
	\bm c_{\ell}^{\top}(\delta) = \bm{H}_{(\ell)} \left( \bm{\mathcal{G}} +  \bm{\mathcal{B}}\bm{\mathcal{M}}(\delta)\right).
\end{equation}

At this point, however, we have only replaced \(\bm{\mathcal{M}}\) with \(\delta\) as decision variables, yet the issue of the nonconvex product of functions remains. Nevertheless, introducing binary indicator variables allows us to reformulate \eqref{eq:analytical-cc-individial-disjoint} as a \textit{disjunctive convex constraint}.

\section{Disjunctive Convex Formulations}
\label{sec:exponential-MIP-formulation}

We propose three approaches to obtain disjunctive convex formulations of \eqref{eq:analytical-cc-individial-disjoint}, which constitute the main contribution of this paper. These approaches yield convex \textit{continuous relaxations} of the mixed-integer optimization problem when solved via B\&B methods.

\begin{remark}
	Obviously, one can solve \(N_{F}\) convex optimization problems by dropping \(\delta\) and using the exhaustive search (ES) method. However, using B\&B allows for exploring the solution space more efficiently, reducing the computational burden and, therefore, finding the optimal solution faster. For further details on B\&B, refer to \cite{Morrison2016}.
\end{remark}

Although motivated by a control application, the approaches presented herein can be generalized to chance constraints involving exclusive disjunctive variables multiplying Gaussian random variables. We simplify notation further by dropping the index \(\ell\) and writing \(\lVert \bm c(\delta)\rVert\) generically as
\begin{equation}
	\lVert \bm c\left(\delta\right) \rVert = \left\lVert \bm g_{0} + \sum_{k=1}^{N_{F}}\delta_{k} \bm g_{k}  \right\rVert
\end{equation}
where $\bm g_{0} = \bm H_{(\ell)}\bm{\mathcal{G}}$ and $\bm g_{k} = \bm H_{(\ell)}\bm{\mathcal{B}}\bm{\mathcal{M}}_{k}$, with $\bm{\mathcal{M}}_{k} = \bm{\mathcal{L}}(\bm{L}_k)\left(\bm I -\bm{\mathcal{B}}\bm{\mathcal{L}}(\bm{L}_k) \right)^{-1}\bm{\mathcal{G}}$.

\subsection{Convex Representations}
\label{sec:convex-reformulations}

To set the stage for what follows, we introduce a property of disjunctive functions essential for the approaches presented here.
\begin{property}
	A function \(h(\cdot)\) acting on mutually exclusive binary indicator variables \(\delta_{k}\), where $\sum_{k=1}^{N_F} \delta_{k}=1$, can be decomposed as
	\begin{equation}
		h\left(\sum \delta_{k}y_{k}\right) = \sum \delta_{k}h\left(y_{k}\right).
		\label{eq:property-disjoint}
	\end{equation}
\end{property}
As a result, \eqref{eq:analytical-cc-individial-disjoint} can be rewritten as
\begin{equation}
	\sum_{k=1}^{N_{F}}\delta_{k}r_{k} \Phi^{-1}(1-\gamma) \leq -f(\mathbf{V})
	\label{eq:analytical-cc-individial-reformulation}
\end{equation}
where
\begin{equation}
	r_{k} = \lVert \bm g_{0} + \bm g_{k} \rVert.
\end{equation}
Recall that we are interested only in the low-risk region (i.e., small values of \(\gamma\)), where \(\Phi^{-1}(1-\gamma) > 0\), which implies that \(f(\mathbf{V}) < 0\).

Furthermore, the formulations that come next give rise to nonconvex function compositions of \(\Phi^{-1}(\cdot)\), which hinders efficiency in finding a solution. To overcome this, we replace them with power- and exponential-cone representable approximations \(\Psi\). This is done using strategies similar to the approach presented in \cite{Barbosa2025}, and we defer details on this until the following subsection.

\subsubsection*{Inverse Probit Approach}
Since \(\delta_{k}\) is binary, \eqref{eq:analytical-cc-individial-reformulation} is equivalent to
\begin{multline}
	\sum_{k=1}^{N_{F}}\delta_{k}^{2}r_{k} \Phi^{-1}(1-\gamma) \leq -f(\mathbf{V}) \iff \\
	\sum_{k=1}^{N_{F}}\left(\delta_{k}\sqrt{r_{k}}\right)^{2} \leq -f(\mathbf{V}) \frac{1}{\Phi^{-1}(1-\gamma)}.
\end{multline}
By introducing a hypograph variable \(t\), we obtain a \textit{rotated second-order cone}\footnote{Defined as the set \(\mathcal{Q}_{\mathrm{r}} = \left\{\left(z,y,t\right) : 2zy \geq t^{2}, z\geq0,y\geq0\right\}\).} constraint
\begin{multline}
	\sum_{k=1}^{N_{F}}\left(\delta_{k}\sqrt{r_{k}}\right)^{2} \leq -f(\mathbf{V}) t \iff \\
	\left\lVert \begin{matrix}
		- t - f(\mathbf{V})\\
		2\sum_{k=1}^{N_{F}}\delta_{k}\sqrt{r_{k}}
	\end{matrix} \right\rVert \leq t - f(\mathbf{V})
\end{multline}
and
\begin{equation}
	t \leq \Psi^{\mathrm{inv}}\left(\gamma\right) \leq \frac{1}{\Phi^{-1}(1-\gamma)},
	\label{eq:constraint-inv}
\end{equation}
where \(1/\Phi^{-1}(1-\gamma)\) is concave on the interval \(\gamma \in [0,0.078]\) and \(\Psi^{\mathrm{inv}}(\gamma)\) its concave, conic-representable lower approximation.

\subsubsection*{Root Probit Approach}
Once again, use the fact that \(\delta_{k}\) is binary and \eqref{eq:analytical-cc-individial-reformulation} is also equivalent to
\begin{multline}
	\Phi^{-1}(1-\gamma) \leq -f(\mathbf{V}) \frac{1}{\sum_{k=1}^{N_{F}}\delta_{k}r_{k} }\iff \\
	\Phi^{-1}(1-\gamma) \leq -f(\mathbf{V})\sum_{k=1}^{N_{F}}\delta_{k}r_{k}^{-1}.
\end{multline}
Then, by introducing an epigraph variable \(t\), we obtain another rotated second-order cone constraint
\begin{multline}
	t^{2} \leq -f(\mathbf{V}) \sum_{k=1}^{N_{F}}\delta_{k}r_{k}^{-1} \iff \\
	\left\lVert \begin{matrix}
		-f(\mathbf{V}) - \sum_{k=1}^{N_{F}}\delta_{k}r_{k}^{-1}\\
		2t
	\end{matrix} \right\rVert \leq -f(\mathbf{V}) + \sum_{k=1}^{N_{F}}\delta_{k}r_{k}^{-1}
\end{multline}
and
\begin{equation}
	\sqrt{\Phi^{-1}(1-\gamma)} \leq \Psi^{\mathrm{root}}\left( \gamma \right) \leq t,
	\label{eq:constraint-root}
\end{equation}
where \(\sqrt{\Phi^{-1}(1-\gamma)}\) is convex on the interval \(\gamma \in [0,0.239]\) and \(\Psi^{\mathrm{root}}(\gamma)\) its convex, conic-representable upper approximation.

\subsubsection*{Logarithm Probit Approach}

By applying monotonic logarithm on both sides of \eqref{eq:analytical-cc-individial-reformulation} and using \eqref{eq:property-disjoint} again, we obtain
\begin{equation}
	\sum_{k=1}^{N_{F}}\delta_{k}\log \left(r_{k} \right) + \log \left( \Phi^{-1}(1-\gamma) \right) \leq \log \left( -f(\mathbf{V}) \right).
\end{equation}
Then, introduce the epigraph variables \(t\) and \(y\), and obtain the convex constraint
\begin{equation}
	\sum_{k=1}^{N_{F}}\delta_{k}\log \left(r_{k} \right) + t \leq y
\end{equation}
with
\begin{align}
	&y \leq \log \left( -f(\mathbf{V}) \right)\text{ and} \label{eq:constraint-otherlog}\\
	&\log{\Phi^{-1}(1-\gamma)} \leq \Psi^{\mathrm{log}}\left(\gamma\right)\leq t,\label{eq:constraint-log}
\end{align}
where \(\log{\Phi^{-1}(1-\gamma)}\) is convex on the interval \(\gamma \in [0,0.158]\) and \(\Psi^{\mathrm{log}}(\gamma)\) its convex, conic-representable upper approximation. It is worth mentioning that \eqref{eq:constraint-otherlog} is exponential-cone representable, as detailed in the following subsection.

In the same way as the probit function, the compositions \(1/\Phi^{-1}(\cdot)\), \(\sqrt{\Phi^{-1}(\cdot)}\) and \(\log{\Phi^{-1}(\cdot)}\) are non-elementary and nonconvex. Therefore, using them directly in the optimization problem results in general nonlinear, nonconvex, and intractable continuous relaxations. On the other hand, replacing them with conic-representable approximations enables convex formulations, allowing the corresponding optimization problems to be solved efficiently.

\subsection{Nonsymmetric Conic Representations}

We obtain guaranteed inner and outer conic-representable approximations of the nonconvex function compositions, which allow \eqref{eq:constraint-inv}, \eqref{eq:constraint-root}, and \eqref{eq:constraint-log} to be represented as conic convex constraints. This is the main idea proposed by Barbosa and Löfberg in \cite{Barbosa2025}, to which we refer for further details.

Thus far, two of the proposed approaches yield constraints that can be represented using the rotated second-order cone, a well-known and widely used symmetric cone. However, to construct the proposed approximations, we make use of two classes of nonsymmetric proper cones: the three-dimensional power and exponential cones. These cones are convex by construction and extend the modeling capabilities beyond what symmetric cones can express, allowing a broader range of convex functions to be represented within the conic optimization framework.

Dahl and Andersen \cite{Dahl2021} generalized the classical primal-dual interior-point algorithm, proposed by Nesterov and Todd \cite{Nesterov1998}, to efficiently solve optimization problems involving nonsymmetric cones. They adapted the primal-dual scalings proposed by Tunçel \cite{Tunel2001} and obtained a more efficient and structured algorithm to handle nonsymmetric cones. Implementations in MOSEK have demonstrated good numerical performance, on level with that of standard symmetric cone algorithms\footnote{The algorithm is specialized for exponential cones.}.

\begin{figure*}[t!] 
	\centering
	\subfloat[]{%
		\includegraphics[width=0.3\textwidth]{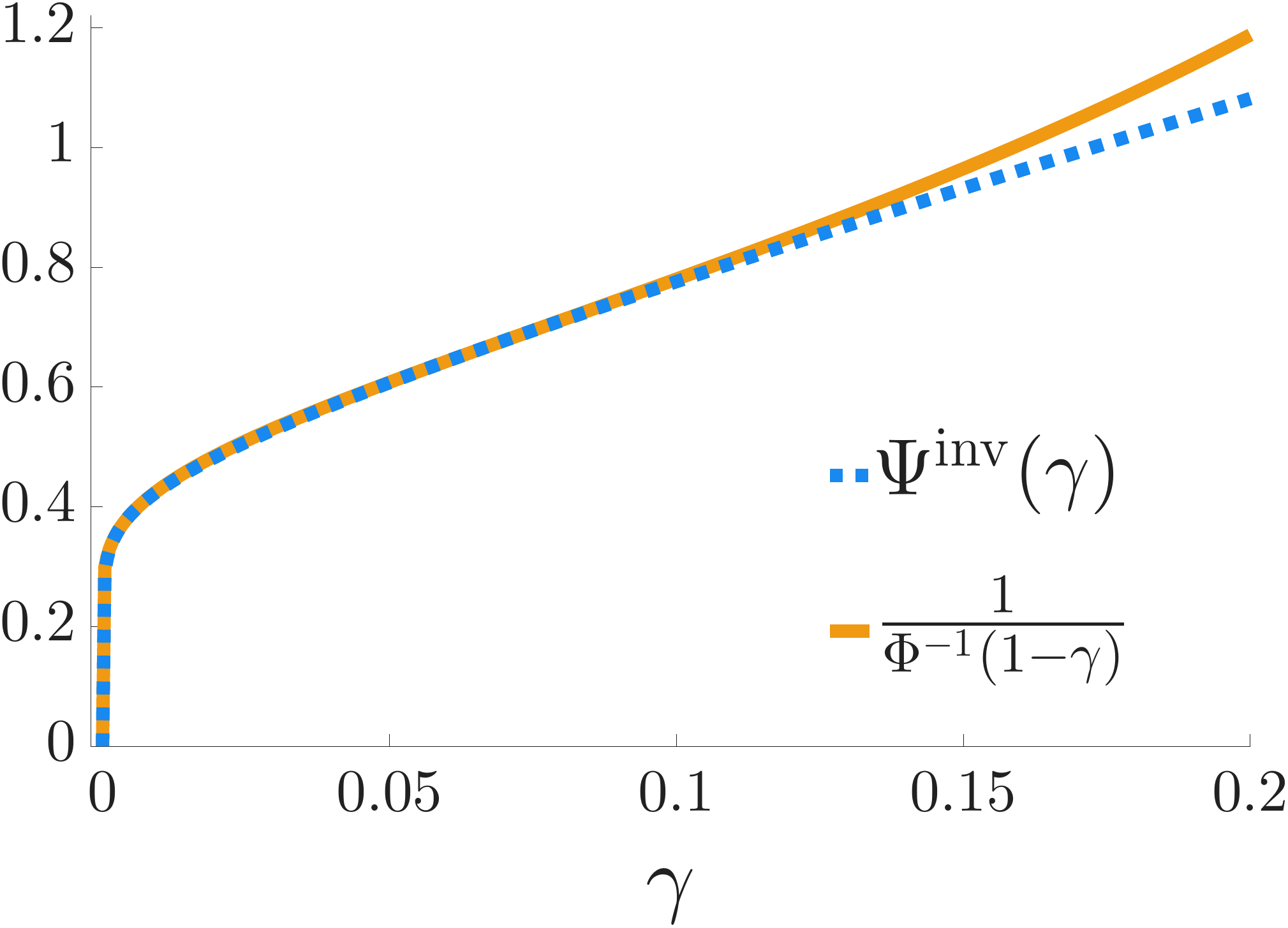}%
		\label{fig:strategy1}
	}
	\hfill
	\subfloat[]{%
		\includegraphics[width=0.3\textwidth]{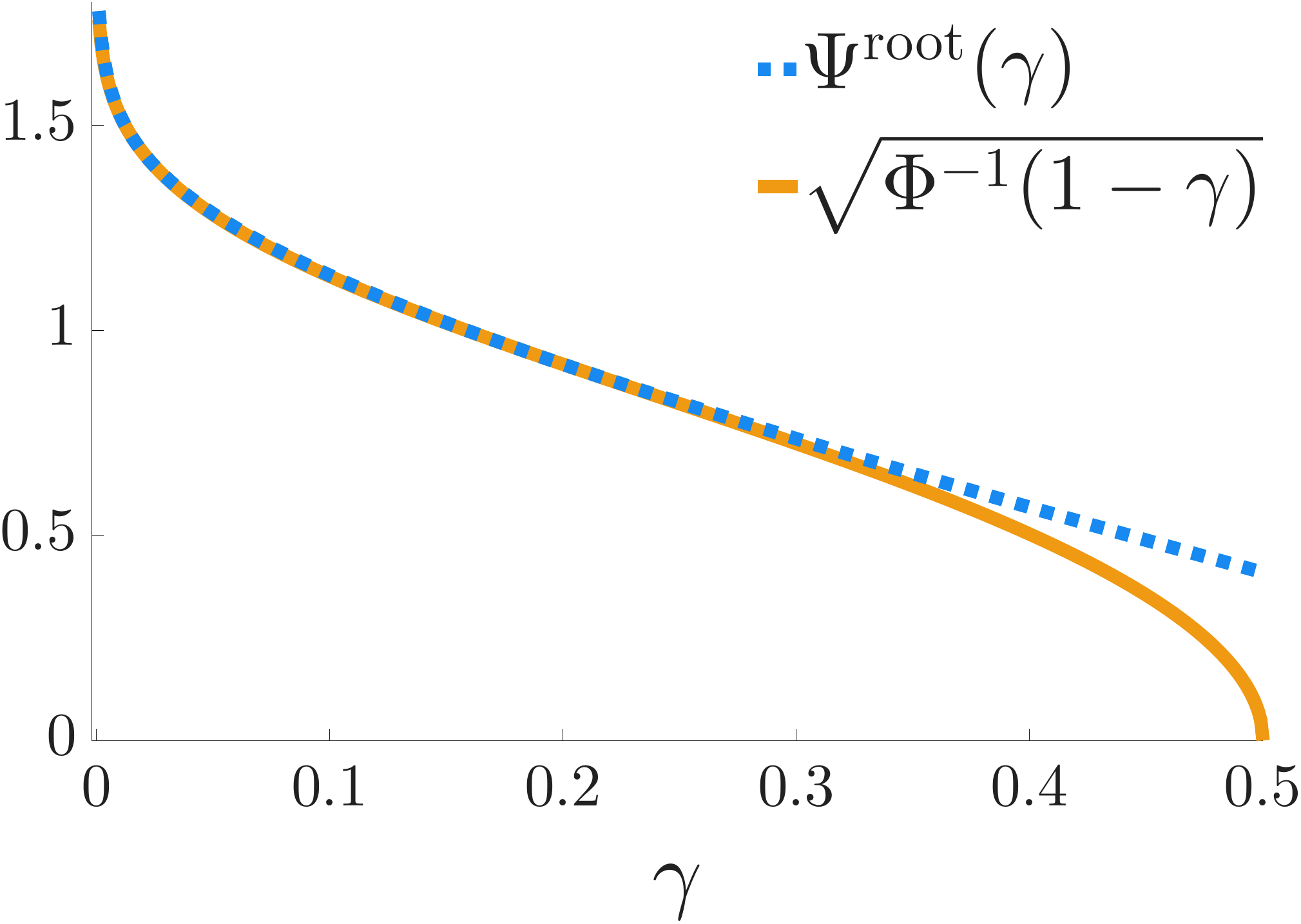}%
		\label{fig:strategy2}
	}
	\hfill
	\subfloat[]{%
		\includegraphics[width=0.3\textwidth]{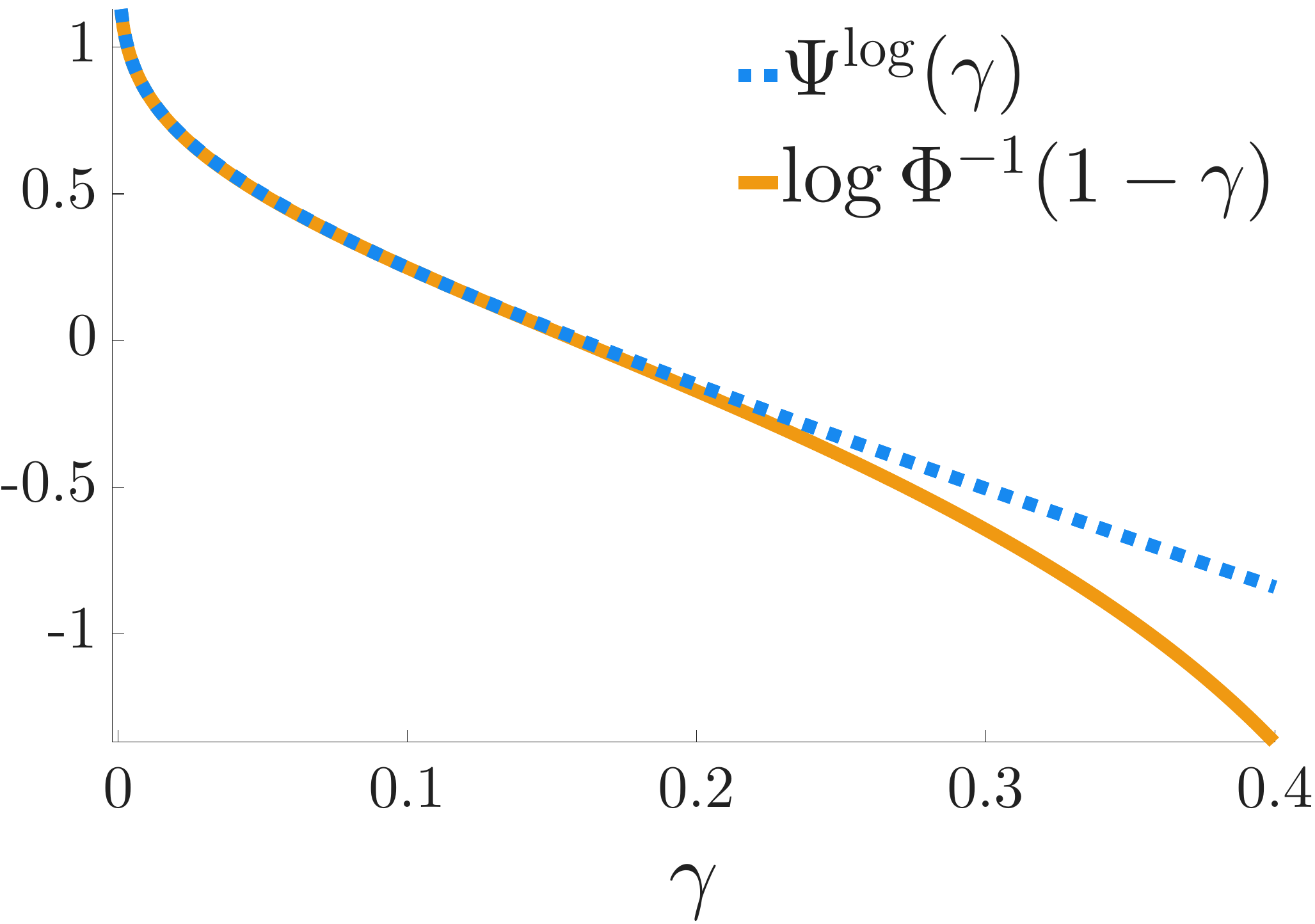}%
		\label{fig:strategy3}
	}
	\caption{The curves of the nonconvex function compositions resulting from the formulations in \autoref{sec:convex-reformulations} (solid) and their corresponding exponential cone-representable approximations (dashed).}
	\label{fig:approximations}
\end{figure*}

\subsubsection{Power cone representation}

The power cone generalizes the classical second-order cone and offers a broader, more flexible structure for formulating problems involving powers other than the specific case of quadratic terms. It is parametrized by \(\eta \in (0,1)\) and defined as
\begin{equation}
	\mathcal{P}_{3}^{\eta,1-\eta} = \left\{(z,y,t) : z^{\eta} y^{1-\eta} \geq |t|,~ z \geq 0,~ y \geq 0\right\}.
	\label{eq:power-cone-def}
\end{equation}

Thus, the power cone enables the representation of convex constraints such as fractional power functions, root-type expressions, inverse power relations, general \(p\)-norms, geometric mean constraints, multiplicative inequalities, inverse product relations, piecewise convex polynomials, and rational bound expressions. For further details, see \cite{Mosek2025}.

In this context, we use the fractional power function as the core component in constructing a compact and accurate approximation of \(1/\Phi^{-1}(1-\gamma)\) in \eqref{eq:constraint-inv}. More specifically, we consider \(z^{\alpha}\) for \(\alpha \in (0,1)\) and \(z\geq 0\), which is concave and can be represented as 
\begin{equation}
	\lvert t \rvert \leq  z^{\alpha} \iff \left(z,1,t\right) \in \mathcal{P}_{3}^{\alpha,1-\alpha}.
\end{equation}

In this way, a lower power cone-representable approximation
\begin{equation}
	\Psi^{\mathrm{inv}}(\gamma) \approx \frac{1}{\Phi^{-1}(1-\gamma)}
	\label{eq:approx-inv}
\end{equation}
can be constructed using a fractional power function combined with compensation terms that preserve convexity and tractability as
\begin{equation}
	\Psi^{\mathrm{inv}} = \beta \gamma^{\alpha} + \lambda \gamma.
\end{equation}
The parameters \(\alpha,\beta\) and \(\lambda\) are determined by solving a standard least-squares curve-fitting problem. The resulting parameter values are presented in \autoref{tab:parameters} and Fig.\ref{fig:strategy1} shows a comparison between the exact function on the right-hand side of \eqref{eq:approx-inv} and its proposed approximation.

\subsubsection{Exponential cone representation}

The exponential cone extends the framework of conic optimization beyond the major polynomial families, i.e., linear, second-order, and power cones, by enabling the incorporation of constraints involving exponential and logarithmic functions. It is defined as
\begin{multline}
	\mathcal{K}_{\mathrm{exp}} = \{(z,y,t) : z \geq ye^{t/y},~ y > 0\}~ \\ 
	\cup ~\{(z,0,t): z\geq 0,~ t \leq 0\}.
	\label{eq:exponential-cone-def}
\end{multline}

From a modeling perspective, the exponential cone can directly represent a wide range of constraints involving exponential and logarithmic functions. More broadly, it enables the representation of convex compositions of functions such as the exponential, logarithm, product logarithm, entropy, relative entropy, softplus, log-sum-exp expressions, and generalized posynomials. For a comprehensive list of functions that can be represented using exponential cones, see \cite{Mosek2025}.

Among these functions, we select the \textit{Lambert W-function} (also known as the product logarithm) as the core component in constructing compact and accurate approximations of \(\sqrt{\Phi^{-1}(1-\gamma)}\) and \(\log\Phi^{-1}(1-\gamma)\). The Lambert W-function is defined as the solution \(W(z)\) that satisfies
\begin{equation}
	z=W(z)e^{W(z)},
	\label{eq:lambertw_standard}
\end{equation}
which is a multivalued function typically defined for complex \(z\) and \(W(z)\). However, for the approximations considered here, we are interested exclusively in its \textit{principal branch} \(W_0:\mathbb{R}_{+} \to \mathbb{R}_{+}\), which is injective and concave. For more details on the Lambert W-function, see \cite{Mez2022}.

Although there is no explicit analytic formula for \(W_0(z)\), the hypograph \(\{(z,y) : 0 \leq z,~0 \leq y \leq W_0(z)\}\) can be described equivalently as
\begin{equation}
	z \geq ye^{y} = ye^{y^{2}/y},
\end{equation}
and modeled as a combination of exponential and second-order cones as
\begin{equation}
	\begin{aligned}
		&(z,y,t) \in \mathcal{K}_{\mathrm{exp}},~(z \geq ye^{t/y})\\
		&(1/2,t,y) \in \mathcal{Q}_{\mathrm{r}},~(t\geq y^2),
	\end{aligned}
\end{equation}
where \(\mathcal{Q}_{\mathrm{r}}\) denotes the rotated second-order cone.

\begin{table}
	\begin{center}
		\caption{Parameter values}
		\label{tab:parameters}
		\begin{tabular}{ l | c | c | c | c |c}
			& \(\alpha\) & \(\beta\) & \(\lambda\) & \(\varphi\) & \(\rho\)\\
			\(\Psi^{\mathrm{inv}}\) & \(0.6406\) & ~~\,\(0.1012\) & ~~\,\(2.6874\) &  & \\
			\(\Psi^{\mathrm{root}}\) & \(2.7465\times10^{3}\) & \(-0.0992\) & \(-1.3059\) & \(1.5798\) & \(-0.0435\)\\
			\(\Psi^{\mathrm{log}}\) & \(3.6416\times10^{2}\) & \(-0.1261\) & \(-2.8898\) & \(0.7186\) & \(-0.0651\)\\
			\hline
		\end{tabular}
	\end{center}
\end{table}

Similarly, the logarithm function, which appeared earlier in \eqref{eq:constraint-otherlog}, is also exponential cone-representable. For \(z \geq 0\), its hypograph can be expressed as
\begin{equation}
	t \leq \log z \iff \left(z,1,t\right)\in\mathcal{K}_{\mathrm{exp}}.
\end{equation}
The logarithm function enters the approximations as a term to address the asymptotic behavior of the probit function, since \(\Phi^{-1}(1-\gamma) \rightarrow \infty\) as \(\gamma \rightarrow 0\). 

In this manner, upper exponential-cone representable approximations 
\begin{align}
	\Psi^{\mathrm{root}}(\gamma) &\approx \sqrt{\Phi^{-1}(1-\gamma)},\label{eq:approx-root}\\
	\Psi^{\mathrm{log}}(\gamma) &\approx \log{\Phi^{-1}(1-\gamma)},\label{eq:approx-log}
\end{align}
can be constructed using \(W_0\) and adding compensation terms that preserve convexity and tractability as
\begin{equation}
	\label{eq:exp_func}
	\Psi(\gamma) = \beta W_{0}(\alpha \,\gamma) + \lambda \gamma + \varphi + \rho \log \gamma.
\end{equation}
Similarly to the previous approximation, the parameters \(\alpha,\beta,\lambda,\varphi\) and \(\rho\) are determined by solving standard least-squares curve fitting problems. \autoref{tab:parameters} presents the resulting values, while Figures \ref{fig:strategy2} and \ref{fig:strategy3} compare the exact functions on right-hand side of \eqref{eq:approx-root} and \eqref{eq:approx-log} with their proposed approximations.

Finally, the nonconvex functions in \eqref{eq:constraint-inv}, \eqref{eq:constraint-root}, and \eqref{eq:constraint-log} can be removed from the constraints and their corresponding power and exponential cone-representable approximations used instead. These approximations enable expressing the problems as mixed-integer cone optimization programs and solving them efficiently.

\section{An Example of Application}


As an example, we consider the problem of selecting a feedback law in the context of chance-constrained optimal path planning with and without obstacles, a typical application of mixed-integer optimization. See, for instance, \cite{Blackmore2011} and \cite{Ioan2021}.
\begin{figure}[t]
	\centering
	\resizebox{\columnwidth}{!}{%
		\begin{tikzpicture}[>=stealth, thick]
		\input{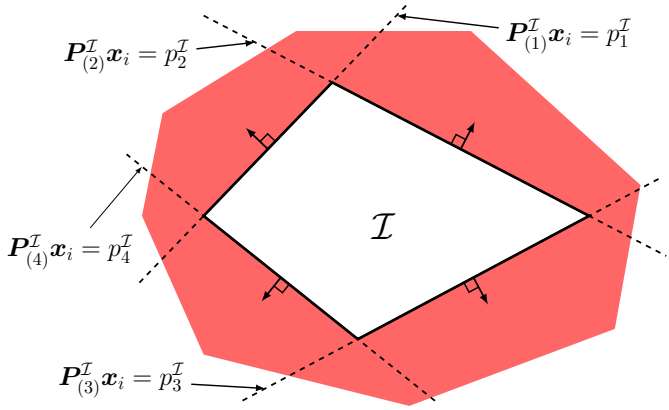}		
		\end{tikzpicture}%
	}
	\caption{A polyhedron representing a stay-in region \(\mathcal{I}\) encoded as a conjunction of linear equality constraints.}
	\label{fig:stay-in}
\end{figure}
\subsection{Problem and Scenario Description}

\begin{figure}[t]
	\centering
	\resizebox{\columnwidth}{!}{%
	\begin{tikzpicture}[>=stealth, thick]
		\input{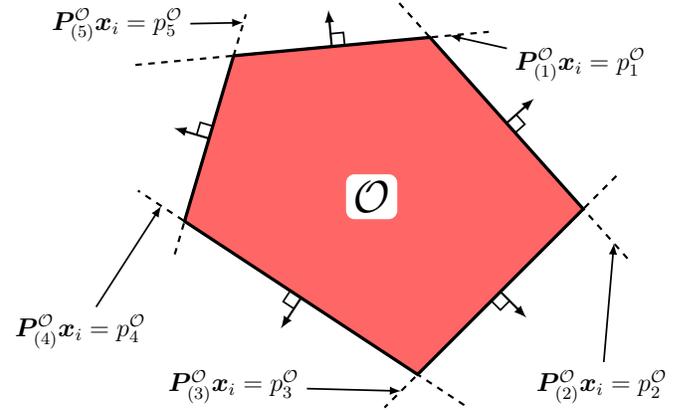}		
	\end{tikzpicture}%
	}
	\caption{A polyhedron representing a stay-out region \(\mathcal{O}\) encoded as a disjunction of linear equality constraints.}
	\label{fig:stay-out}
\end{figure}

Consider the path-planning problem in which an unmanned vehicle must remain within a designated \textit{stay-in} region \(\mathcal{I}\) and outside a \textit{stay-out} region \(\mathcal{O}\), both defined as polyhedral convex sets. The \textit{stay-in} region is represented as a conjunction of linear constraints as
\begin{equation}
	\mathcal{I} \iff \bigwedge_{i=1}^{N} \bigwedge_{\ell=1}^{N_{\mathcal{I}}} \bm{P}_{(\ell)}^{\mathcal{I}} \bm x_{i} \leq p_{\ell}^{\mathcal{I}},
	\label{eq:stay-in}
\end{equation}
where \(N_{\mathcal{I}} \in \mathbb{N}\) is the number of faces of  \(\mathcal{I}\). The \textit{stay-out} region is represented as a disjunction of linear constraints as
\begin{equation}
	\mathcal{O} \iff \bigwedge_{i=1}^{N} \bigvee_{\ell=1}^{N_{\mathcal{O}}}\bm{P}^{\mathcal{O}}_{(\ell)} \bm x_{i} \geq p_{\ell}^{\mathcal{O}},
	\label{eq:stay-out}
\end{equation}
where \(N_{\mathcal{O}} \in \mathbb{N}\) is the number of faces of  \(\mathcal{O}\). Illustrative examples of these regions are depicted in Figures \ref{fig:stay-in} and \ref{fig:stay-out}.

Thus, in this application, the vehicle must remain within \(\mathcal{I}\) and outside \(\mathcal{O}\). To this end, mutually exclusive binary variables are used to model both the selection of the feedback law and the enforcement of the stay-out region constraints.

\subsection{Feedback Selection}

At each sampling instant, one feedback law \(\bm L_{k}\) is selected to be used over the entire prediction horizon, while the vehicle remains within \(\mathcal{I}\) at every predicted step with probability at least \(1-\gamma^{\mathcal{I}}\). This is enforced through individual chance constraints as
\begin{equation}
	\bigwedge_{i=1}^{N} \bigwedge_{\ell = 1}^{N_{\mathcal{I}}} \bigvee_{k = 1}^{N_{F}} \mathbb{P}\left(\bm{P}_{(\ell)}^{\mathcal{I}} \bm x_{i+1}(\bm v_{i}, \delta_{k}) \leq p_{\ell}^{\mathcal{I}}\right) \geq 1 - \gamma_{i + 1}^{\mathcal{I}}.
	\label{eq:chance-constraint-stay-in}	
\end{equation}
Additionally, since the control inputs depend on stochastic variables, their constraints must also be probabilistic. Thus, \(\bm u_{i}\) must remain within the polyhedral convex input constraint set \(\mathcal{U}\) with probability at least \(1-\gamma^{\mathcal{U}}\) as
\begin{equation}
	\bigwedge_{i=1}^{N} \bigwedge_{\ell=1}^{N_{\mathcal{U}}} \bigvee_{k = 1}^{N_{F}} \mathbb{P}\left(\bm{P}_{(\ell)}^{\mathcal{U}} \bm u_{i}(\bm v_{i}, \delta_{k}) \leq p_{\ell}^{\mathcal{U}}\right)\geq 1 - \gamma_{i}^{\mathcal{U}},
	\label{eq:chance-constraint-input}
\end{equation}
where \(\mathcal{U}\) is defined in the same manner as \(\mathcal{I}\). 

Moreover, the selection of a single feedback law throughout the prediction horizon is enforced by
\begin{equation}
	\sum_{k=1}^{N_F} \delta_{k} = 1.
	\label{eq:only-one-feedback}
\end{equation}
Note that the constraints in \eqref{eq:chance-constraint-stay-in} and \eqref{eq:chance-constraint-input} evaluate in the same fashion as \eqref{eq:chance-selection}--\eqref{eq:analytical-cc-individial-reformulation}.

\subsection{Avoiding the Stay-out Region}

At each prediction step \(i\), the vehicle must remain outside \(\mathcal{O}\) with probability at least \(1- \gamma^{\mathcal{O}}\). Enforcing obstacle avoidance introduces an additional layer of disjunction into the chance constraint formulation
\begin{equation}
	\bigwedge_{i=1}^{N} \bigvee_{\ell=1}^{N_{\mathcal{O}}} \bigvee_{k = 1}^{N_{F}} \mathbb{P}(\bm{P}_{(\ell)}^{\mathcal{O}} \bm x_{i+1}(\bm v_{i}, \delta_{k})\!\geq\! p_{\ell}^{\mathcal{O}})\geq 1 - \gamma_{i + 1}^{\mathcal{O}}.
	\label{eq:chance-constraint-stay-out}
\end{equation}

The condition ensuring that the vehicle's position lies outside \(\mathcal{O}\) can be modeled using \textit{Big-M} constraints as 
\begin{gather}
	\mathbb{P} \left(\bm{P}_{(\ell)}^{\mathcal{O}} \bm x_{i+1}(\bm v_{i}, \delta_{k}) \!\geq\! p_{\ell}^{\mathcal{O}} - \mathfrak{M}\left(1 - \sigma_{i\,\ell}\right)\right) \!\geq\! 1 - \gamma_{i+1}^{\mathcal{O}}\label{eq:constraint-big-M},\\
	\sum_{\ell=1}^{N_{\mathcal{O}}} \sigma_{i \, \ell} = 1, \label{eq:enforce-one-avoidance}
\end{gather}
where \(\sigma \in \{1,0\}\) and \(\mathfrak{M}\) is chosen sufficiently large so that the \(\ell\)-th constraint in \eqref{eq:constraint-big-M} is active when \(\sigma_{i\,\ell} = 1\) and trivially satisfied otherwise. The constraint \eqref{eq:enforce-one-avoidance} guarantees exactly one active avoidance constraint per prediction step. 

The \textit{Big-M} method is a classical technique widely used in MIO problems, including path planning with obstacle avoidance. See, for instance, \cite{Ioan2021} and \cite{Schouwenaars2001}. Moreover, it is important to note that adding this selection does not compromise the formulations presented in \autoref{sec:exponential-MIP-formulation}. See Appendix \ref{sec:appendix1}.

\section{Numerical Example}

The application described above (chance-constrained optimal path planning with avoidance regions) is now demonstrated through a simple example. An SMPC is used to control an unmanned vehicle operating in a two-dimensional region under disturbances. The vehicle must navigate from a given initial state to inside a designated \textit{target region} \(\mathcal{T}\). Two scenarios are considered: without obstacles (Case 1) and with a single obstacle (Case 2), with feedback law selection occurring in both. The OCP is modeled using the YALMIP toolbox \cite{Lofberg2004} and solved with MOSEK.

\subsection{Setup}

The state and control inputs vectors of the vehicle are defined respectively as
\begin{equation}
	\bm x_{i} = 
	\begin{bmatrix}
		v^{x}_{i} & p^{x}_{i} & v^{y}_{i} & p^{y}_{i}
	\end{bmatrix}^{\top} \text{and } \bm u_{i} = 
	\begin{bmatrix}
		u^{x}_{i}\\
		u^{y}_{i}
	\end{bmatrix},
\end{equation}
where \(p^{x}_{i}\) and \(p^{y}_{i}\) are the vehicle's positions, and \(v^{x}_{i}\) and \(v^{y}_{i}\) the vehicle's velocities along the \(x\)- and \(y\)-axis. Its discrete-time dynamics are given by
\begin{multline}
	\bm A = 
	\begin{bmatrix}
		1 & 0 & 0 & 0\\
		0.1 & 1 & 0 & 0\\
		0 & 0 & 1 & 0\\
		0 & 0 & 0.1 & 1
	\end{bmatrix}, 
	\bm B = 
	\begin{bmatrix}
		0.1 & 0\\
		0.005 & 0\\
		0 & 0.1\\
		0 & 0.005		
	\end{bmatrix}, \\ \text{and } \bm G = 
	\begin{bmatrix}
		0.1 & 0\\
		0.005 & 0\\
		0 & 0.1\\
		0 & 0.005
	\end{bmatrix}. \qquad \qquad \quad \quad
\end{multline}
and the feasible input set is defined by the polyhedron
\begin{equation}
	\mathcal{U} = \left\{ \bm u :
	\begin{bmatrix}
		-5\\
		-5
	\end{bmatrix} \leq \bm u \leq
	\begin{bmatrix}
		5\\
		5
	\end{bmatrix}\right\}.
\end{equation} 
The target region is modeled as another stay-in region satisfying \(\mathcal{T} \cap \mathcal{I}\). To ensure that the vehicle reaches it, we require the terminal positions \(p^{x}_{N+1}\) and \(p^{y}_{N+1}\) to lie within \(\mathcal{T}\) with probability at least \(1-\gamma^{\mathcal{T}}\), enforced as
\begin{equation}
	\bigwedge_{\ell=1}^{N_{\mathcal{T}}} \bigvee_{k = 1}^{N_{F}} \mathbb{P}\left(\bm{P}_{(\ell)}^{\mathcal{T}} \bm x_{N+1}(\bm u_{N}, \delta_{k}) \leq p_{\ell}^{\mathcal{T}}\right)\geq 1 - \gamma_{N+1}^{\mathcal{T}}.
	\label{eq:terminal-constraint}
\end{equation}

In case 1, the risk allocation associated with the stay-in region is treated as decision variables and stacked as
\begin{equation*}
	\bm \Gamma = \left[\gamma_{1}^{\mathcal{I}},\dots,\gamma_{N}^{\mathcal{I}}\right]^\top.
\end{equation*}
In Case 2, both the risks for the stay-in and stay-out regions are considered decision variables, stacked as
\begin{equation*}
	\bm \Gamma =
	\begin{bmatrix}
		\gamma_{1}^{\mathcal{I}} & \dots & \gamma_{N}^{\mathcal{I}},
		\gamma_{1}^{\mathcal{O}} & \dots & \gamma_{N}^{\mathcal{O}}
	\end{bmatrix}^\top.
\end{equation*}

For convenience, the risk allocations associated with the chance constraints on the control inputs and terminal positions are fixed as \(\gamma_{i}^{\mathcal{U}} = \gamma_{i}^{\mathcal{T}} = 10^{-2}\). We believe that this simplification does not compromise generality, as the risk allocations for the stay-in and stay-out regions (when applicable) remain decision variables.

The pre-computed feedback laws are given by \(N_{F}\) discrete-time linear-quadratic controllers. Each gain \(\bm L_{k}\) is uniquely determined by a combination of the parameters in the weighting matrices
\begin{equation}
	\bm Q^{F} = \mathrm{diag}\left(0,r_{m},0,1\right) \text{ and } \bm R^{F} = \mathrm{diag}\left(r_{n},r_{p}\right),
	\label{eq:weighting-feedbacks}
\end{equation}
where \((m, n, p) \in \{1, \dots, N_{L}\}^3\), \(N_{L}\in \mathbb{N}\) and \(r\) is uniformly sampled within a given interval. Each unique triplet \((m, n, p)\) defines a distinct \(\bm L_{k}\), yielding \(N_{F} = N_{L}^3\) feedback laws. The parameter ranges are defined as \(r_{s} \in \left[r^{\mathrm{min}},r^{\mathrm{max}}\right]\) \(\forall s \in \{1,\dots,N_{L}\}\). Subsequently, \(\bm{\mathcal{M}}(\delta)\) is computed according to \eqref{eq:disturbance-feedback-equivalence-selection}, covering the entire grid of possible linear-quadratic controller designs over the specified parameter sets.

The cost function \(\mathrm{J}\) trades off the risk of chance constraint violations and the cost of using the selected feedback law over the prediction horizon. Due to the stochastic nature of the control inputs, the expected value of the cost function is minimized
\begin{multline}
	\mathbb{E}\left[\mathrm{J}\right] = \mathbb{E} \left[ \bm{\mathcal{S}}^{\top} \bm\Gamma + \sum_{k = 1}^{N_{F}}\lVert \mathbf{U}(\mathbf{V},\delta_{k})\rVert_{\bm{\mathcal{R}}}^{2} \right] \\ =
	\bm{\mathcal{S}}^{\top} \bm\Gamma + \mathbf{V}^{\top} \bm{\mathcal{R}} \mathbf{V} + \sum_{k=1}^{N_{F}} \mathrm{tr} \left(\!\bm{\mathcal{M}}^{\top}\!\left(\delta_{k}\right) \bm{\mathcal{R}} \bm{\mathcal{M}}\left(\delta_{k}\right)\right)\!\delta_{k}
	\label{eq:cost}
\end{multline}
where \(\bm{\mathcal{R}} = \oplus_{i=1}^{N} \bm R\) and \(\bm{\mathcal{S}} = \bm{1}_{N} \otimes \bm{S}\) with the appropriate dimensions.  The matrices in \eqref{eq:weighting-feedbacks} are solely used to compute \(\bm{\mathcal{M}}(\delta)\); the control inputs in \(\mathrm{J}\) are penalized using fixed \(\bm R = \mathrm{diag}(0.05,0.05)\). See Appendix \ref{sec:appendix2} for details on expected quadratic costs involving Gaussian random variables.

The general formulation of the SMPC problem, applied in both cases presented below is
\begin{equation}
	\begin{aligned}
		& \underset{\mathbf{V}, \delta}{\mathrm{minimize}}
		& & \mathbb{E} \left[ \mathrm{J} \right] \\
		& \mathrm{subject~to}
		& & \text{\eqref{eq:state-space-closed-selection}, \eqref{eq:chance-constraint-stay-in},\eqref{eq:chance-constraint-input},\eqref{eq:only-one-feedback} and \eqref{eq:terminal-constraint}} \\
		& & & 0 \leq \bm{1}^{\top}\bm{\Gamma} \leq \xi
	\end{aligned}
\end{equation}
with \(\xi = 0.15\), \(N = 10\), \( N_{\mathcal{I}}=4\), \(N_{\mathcal{T}} = 2\) and \(N_{F} = 125\)  (i.e., \(N_{L}=5\)). The chance constraints evaluate as described in \eqref{eq:chance-selection}--\eqref{eq:analytical-cc-individial-reformulation}. Following this, one of the formulations presented in \autoref{sec:convex-reformulations}, along with the corresponding approximation \eqref{eq:approx-inv}, \eqref{eq:approx-root}, or \eqref{eq:approx-log}, is then selected to obtain a convex deterministic representation of the chance constraints. Finally, the resulting SMPC problem is solved as a mixed-integer conic optimization problem.

\subsection{Results and Discussion}

\begin{figure}[!t]
	\centering
	\subfloat[Best and worst feasible integer solution using \text{\(r_{s} \in \left[0.0005,0.3\right]\)}]{
		\includegraphics[width=\columnwidth]{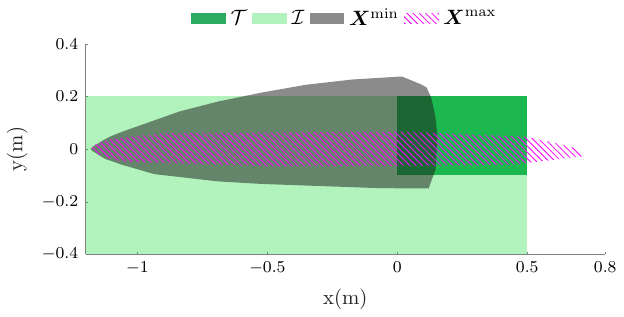}
		\label{fig:feedback-selection-0005}
	}
	\hfill
	\centering
	\subfloat[Best and worst feasible integer solution using \text{\(r_{s} \in \left[0.005,0.3\right]\)}]{
		\includegraphics[width=\columnwidth]{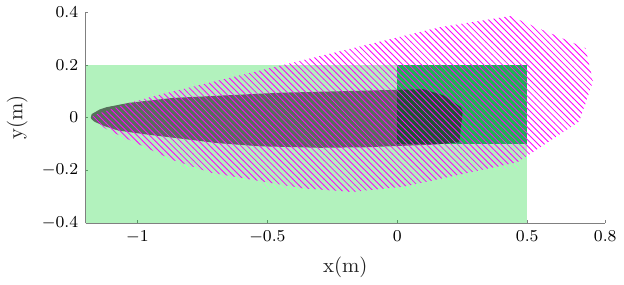}
		\label{fig:feedback-selection-005}
	}
	\caption{1000 Monte Carlo simulations of the state predictions at the first sampling instant, obtained using the best and worst feasible integer solutions. The plots show the prediction envelopes of the best integer solution \(\mathbf{X}^{\mathrm{min}}\) (gray) and the worst integer solution \(\mathbf{X}^{\mathrm{max}}\) (magenta). The results in each subplot correspond to different parameter ranges used to construct \(\bm{\mathcal{M}}(\delta)\).}
	\label{fig:feedback-slection-open-loop}
\end{figure}

Considering the scenario in Case 1 and \(\bm S=2\),  \autoref{fig:feedback-slection-open-loop} shows the state prediction envelopes obtained by applying the control sequences from the best and worst performing feasible integer solutions (with respect to \(\mathbb{E}[\mathrm{J}]\)) at the very first sampling instant, denoted by \(\mathbf{U}^{\mathrm{min}}\) and \(\mathbf{U}^{\mathrm{max}}\), respectively. The corresponding state envelopes \(\mathbf{X}^{\mathrm{min}}\) and \(\mathbf{X}^{\mathrm{max}}\) are generated from \(1000\) Monte Carlo simulations each.

The results in Fig.~\ref{fig:feedback-selection-0005} were obtained for \(r_{s} \in \left[0.0005,0.3\right]\). In this case, the envelope of \(\mathbf{X}^{\mathrm{min}}\) was obtained with \(\bm{\mathcal{M}}(\delta_{56})\), constructed with
\begin{equation*}
	\bm L_{56} = 
	\begin{bmatrix}
		-5.0848 & -12.9277 &  0  & 0\\
		0 & 0 & -2.5231 & -3.1829
	\end{bmatrix}.
\end{equation*}
Conversely, the envelope of \(\mathbf{X}^{\mathrm{max}}\) was obtained with \(\bm{\mathcal{M}}(\delta_{30})\), constructed with
\begin{equation*}
	\bm L_{30} = 
	\begin{bmatrix}
		-0.9765 & -0.4768 & 0 & 0\\
		0 & 0 & -7.4821 & -27.9909
	\end{bmatrix}.
\end{equation*} 

The results in Fig.~\ref{fig:feedback-selection-005} were obtained for \(r_{s} \in \left[0.005,0.3\right]\). Here, the envelope of \(\mathbf{X}^{\mathrm{min}}\) was obtained with \(\bm{\mathcal{M}}(\delta_{101})\), constructed with
\begin{equation*}
	\bm L_{101} = 
	\begin{bmatrix}
		-3.5677 & -6.3642 & 0 & 0\\
		0 & 0 & -4.6580 & -10.8484
	\end{bmatrix}.
\end{equation*}
Conversely, the envelope of \(\mathbf{X}^{\mathrm{max}}\) was obtained with \(\bm{\mathcal{M}}(\delta_{12})\), constructed with
\begin{equation*}
	\bm L_{12} = 
	\begin{bmatrix}
		-0.6974 & -0.2432 & 0 & 0\\
		0 & 0 & -2.1386 & -2.2869
	\end{bmatrix}.
\end{equation*}

The best-performing solutions have consistently higher feedback gains associated with the \(x\)-direction compared to the worst-performing ones. However, in \(\bm L_{56}\), the feedback gains in the \(y\)-direction are insufficient. As a result, several predicted terminal positions violate the boundaries of \(\mathcal{T}\), as shown in Fig.~\ref{fig:feedback-selection-0005}. By contrast, \(\bm L_{101}\), has increased \(y\)-direction gains, ensuring that the predicted terminal positions remain within \(\mathcal{T}\), as shown in Fig.\ref{fig:feedback-selection-005}.

The poor performance with \(\bm L_{30}\), shown in Fig.~\ref{fig:feedback-selection-0005}, is the result of excessively high gains in the \(y\)-direction, which increase the cost. However, it might be counter intuitive that \(\bm L_{12}\), despite having the smallest gains in all of its entries among the analyzed state feedback matrices, yields the highest cost and poorest overall state performance. See Fig.\ref{fig:feedback-selection-005}. The reason for this is that lower feedback gains in \(\bm L\) directly translate into weaker disturbance feedback in \(\bm{\mathcal{M}}\) i.e., approaching an open-loop control sequence (recall \eqref{eq:input-disturbance-feedback}). This reduces disturbance rejection and leads to larger values of the nominal control input sequence \(\mathbf{V}\).

\begin{figure}[!t]
	\centering
	\includegraphics[width=\columnwidth]{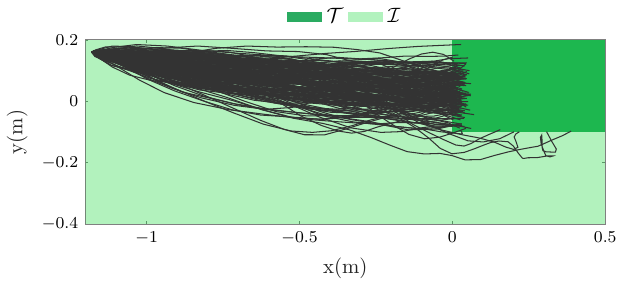}
	\caption{100 Monte Carlo simulations of the performed trajectories in Case 1 (without obstacles). The  vehicle navigates inside $\mathcal{I}$ (light green), from the initial state \(\bm x_{0} = \left[0,-1.18,0,0.16\right]^{\top}\), to $\mathcal{T}$ (dark green). As the vehicle's initial position is close to the boundaries of \(\mathcal{I}\), a safer trajectory toward \(\mathcal{T}\) involves moving away from these limits.}
	\label{fig:example-scenario-no-obstacles}
\end{figure}
\begin{figure}[t!]
	\centering
	\includegraphics[width=\columnwidth]{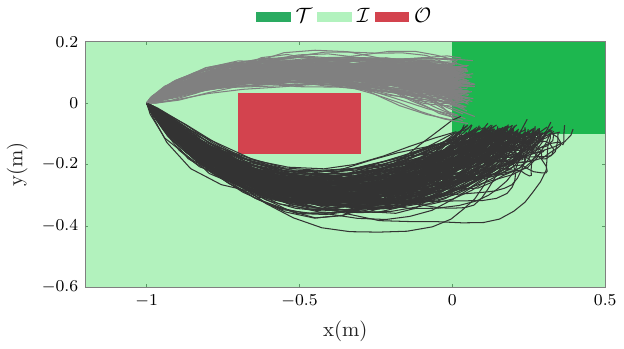}
	\caption{100 Monte Carlo simulations of the performed trajectories in Case 2 (with obstacles). The vehicle navigates within $\mathcal{I}$ (light green), from the initial state \(\bm x_{0} = \left[0,-1,0,0\right]^{\top}\), toward $\mathcal{T}$ (dark green), while avoiding $\mathcal{O}$ (red). Larger weights on the risk allocation (\(S = \mathrm{diag}\left(10,10\right)\)) produce the trajectories shown with dark lines. Lower weights on the risk allocation (\(S = \mathrm{diag}\left(1,1\right)\)) produce the trajectories shown with gray lines.}
	\label{fig:example-scenario-obstacle}
\end{figure}

Figures~\ref{fig:example-scenario-no-obstacles} and \ref{fig:example-scenario-obstacle} show 100 Monte Carlo simulations of the performed trajectories obtained for Case 1 and 2, respectively. In Case 1, we consider \(r_{s} \in \left[0.05,0.3\right]\) and \(\bm S = 2\); in Case 2 we consider \(r_{s} \in \left[0.1,0.15\right]\) and either \(\bm S = \mathrm{diag}\left(1,1\right)\) or \(\bm S = \mathrm{diag}\left(10,10\right)\).

In Case 1, one observes the back-off in the average performed trajectory, caused by the risk allocation. Since the vehicle's initial position is close to the boundaries of \(\mathcal{I}\), a safer trajectory toward \(\mathcal{T}\) involves moving away from these limits. More interestingly, in Case 2, the choice of \(\bm S\) in weighting the risk allocation \(\bm \Gamma\) significantly influences how the vehicle avoids the stay-out region \(\mathcal{O}\). Lower weights \(\bm S = \mathrm{diag}\left(1,1\right)\) lead to shorter, less conservative trajectories. On the other hand, increasing the weights on the risk allocation to \(\bm S = \mathrm{diag}\left(10,10\right)\) implies that we increased the importance of keeping a safe distance from the boundaries of \(\mathcal{I}\) and \(\mathcal{O}\), which consequently leads to longer, conservative trajectories.

\begin{figure}[!t]
	\centering
	\subfloat[The time to compute a solution at every sampling time in Case 1.]{
		\includegraphics[scale=0.2]{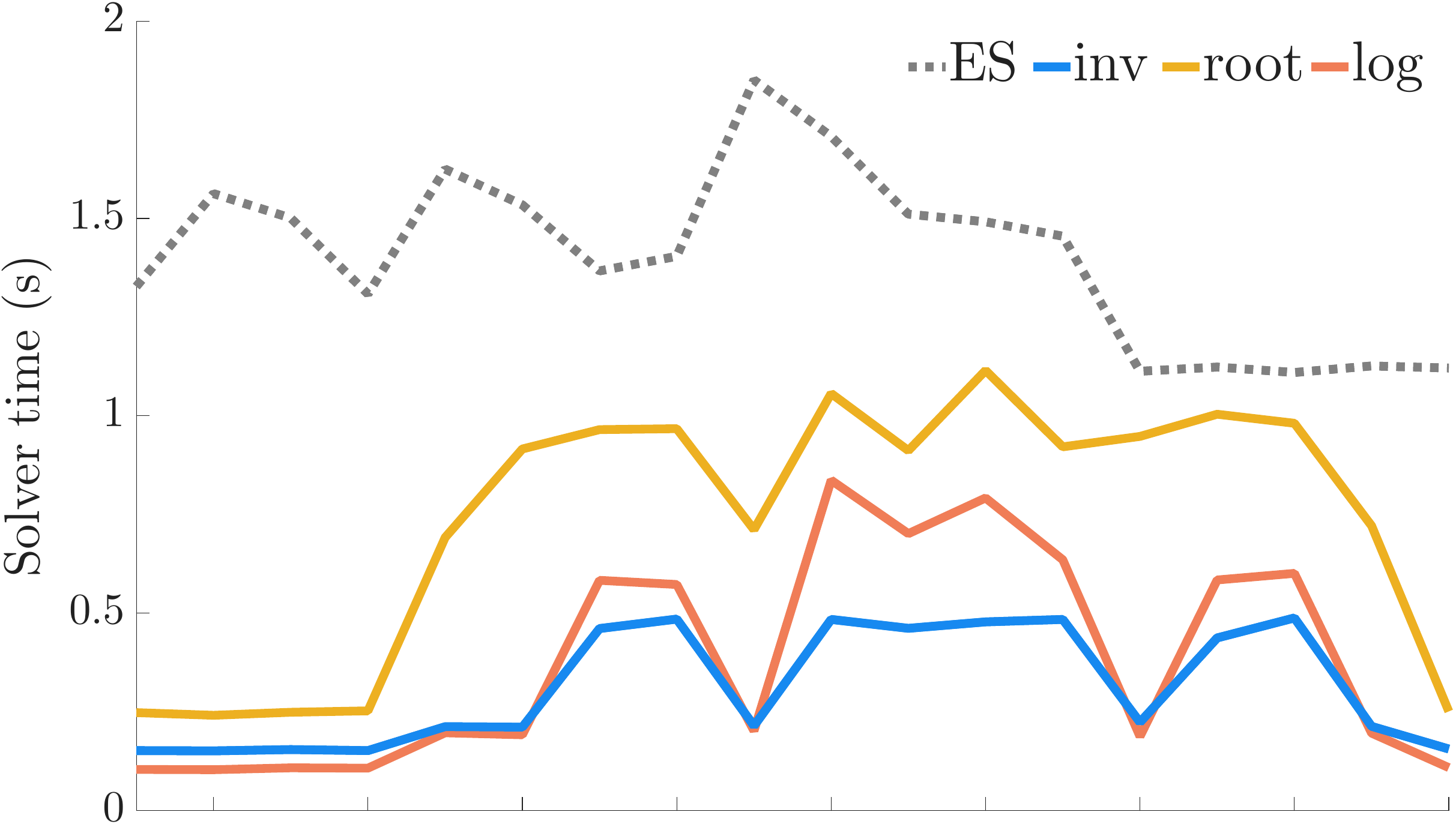}
		\label{fig:time-no-obstacle}
	}\hfill
	\subfloat[The time to compute a solution at every sampling time in Case 2.]{
		\includegraphics[scale=0.2]{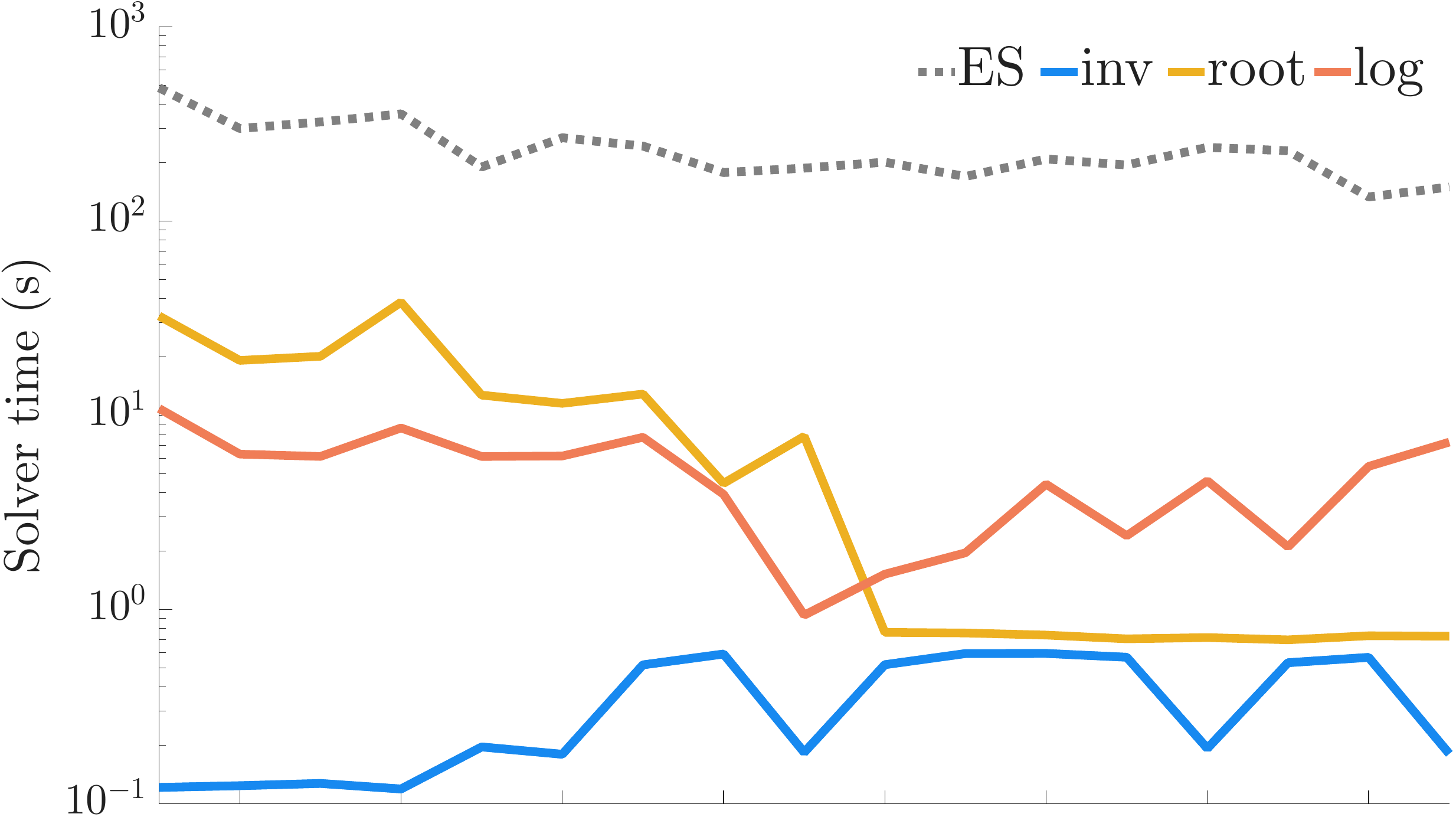}	\label{fig:time-obstacle}
	}
	\caption{The time to solve the optimization problem at every sampling time in both cases. The solution times of the Inverse (blue), Root  (yellow), and Logarithm (orange) probit approaches are compared to the exhaustive search method (dashed gray). Note the logarithm scale for the solution times in Case 2.}
	\label{fig:times}
\end{figure}

\autoref{fig:times} compares the solver times per sampling instant for a single trajectory realization, using the proposed approaches and the ES method\footnote{The ES method uses an updated version of the approximation from \cite{Barbosa2025} that addresses the asymptotic behavior of \(\Phi^{-1}\).} in both cases. The comparison uses the same disturbance realization, generated with a Mersenne Twister (seed \(3\)). The setup is the same as used to obtain the trajectory realizations above, and the solver times for Case 2 were obtained for the longer, more conservative trajectory, i.e., \(\bm S = \mathrm{diag}\left(10,10\right)\). The mean and standard deviation of the solver times are reported in \autoref{tab:comparison-mean-st}. The trajectory in Case 1 involved \(5\) feedback choices, while the ones in Case 2 involved \(6\). However, it is important to mention that the feedback selection can be different for other disturbance realizations.

In both cases, the optimization problems are more efficiently solved using the proposed approaches than using the ES method. Although already noticeable in Fig.~\ref{fig:time-no-obstacle}, this advantage is further highlighted in Fig.~\ref{fig:time-obstacle} (note the logarithmic scale). Furthermore, the same can be concluded upon inspection of \autoref{tab:comparison-mean-st}.

Lastly, it is important to mention that since the approximations presented here are conservative, any bound on the risk allocation \(\xi\in[0,0.5]\) can be used. However, the approximations degrade and become increasingly conservative for larger \(\xi\). It is also worth noting that recursive feasibility was not addressed. 

\begin{table}[t!]
	\centering
	\caption{Mean and standard deviation (s.d.) of the solution times (s).}
	\label{tab:comparison-mean-st}
	\begin{tabular}{l|c|c|c|c|c}
		\multicolumn{2}{c|}{} & \(\mathrm{inv}\) & \(\mathrm{root}\) & \(\mathrm{log}\) & ES\\
		\hline 
		\multirow{2}{*}{\shortstack[l]{Case 1}} 
		& mean & \(0.3124\) & \(0.7309\) & \(0.3787\) & \(1.4022\) \\
		& s.d. & \(0.1498\) & \(0.3270\) & \(0.2724\) & \(0.2219\)\\
		\hline
		\multirow{2}{*}{\shortstack[l]{Case 2}} 
		& mean & \(0.3467\) & \(9.2008\) & \(4.9042\) & \(224.0117\)\\
		& s.d. & \(0.2090\) & \(11.6327\) & \(2.7575\) & \(88.4119\)\\
	\end{tabular}
\end{table}

\section{Conclusions}

This paper presented three approaches to derive disjunctive convex constraints for SMPC problems where the feedback law and risk allocation are optimized over. These formulations yield convex continuous relaxations of the OCP, enhancing computational efficiency when using B\&B algorithms. The main advantage is that the problem can be formulated as mixed-integer conic optimization, for which structured solvers and established algorithmic frameworks are available. 

The proposed formulations were validated through an SMPC application in a path planning problem under additive disturbances. Furthermore, they can be generalized to chance constraints involving mutually exclusive binary variables multiplying Gaussian stochastic variables, that is, optimization problems that involve selecting a single linear combination of stochastic variables in the chance constraints. Future work will focus on incorporating Gaussian mixture random variables into the framework.

\appendices
\section{Implication of Big-M Formulation}
\label{sec:appendix1}
A chance constraint with disjunctive variables multiplying the stochastic variables can be turned on and off via the Big-M method as
\begin{equation}
	\mathbb{P}\left( f (\mathbf{V}) + \bm c^{\top}(\delta) \mathbf{W} \leq  \mathfrak{M}\left(1 - \sigma\right)\right) \geq 1-\gamma
	\label{eq:chance-bigM}
\end{equation}
which evaluates to
\begin{equation}
	f(\mathbf{V}) + \lVert \bm c(\delta)\rVert\Phi^{-1}(1-\gamma) \leq  \mathfrak{M}\left(1 - \sigma\right)
	\label{eq:analytical-cc-bigM}
\end{equation}
and the convex formulations presented in \autoref{sec:convex-reformulations} can be applied.

\section{Analytical Form of the Cost Function}
\label{sec:appendix2}

To formulate a quadratic cost function involving stochastic Gaussian variables in the analytical form, we use the following property from \cite{Petersen2008}.
\begin{property}
	\label{sec:property2}
	Assume \(\bm \Upsilon\) is symmetric and let \(\bm \Sigma = \mathrm{Var}\left[\mathbf{W}\right]\), then
	\begin{equation}
		\mathbb{E}\left[\mathbf{W}^{\top} \bm \Upsilon \mathbf{W}\right] = \mathrm{tr}\left(\bm \Upsilon \bm \Sigma \right) + \mathbb{E}\left[\mathbf{W}^{\top}\right] \bm \Upsilon \mathbb{E}\left[\mathbf{W}\right].
	\end{equation}
\end{property}

A quadratic cost function with stochastic variables on the states and inputs can be written as
\begin{equation}
	\mathbb{E}\left[\lVert \mathbf{X} \rVert_{\bm{\mathcal{Q}}}^{2} + \lVert \mathbf{U} \rVert_{\bm{\mathcal{R}}}^{2}\right] = \mathbb{E}\left[\mathbf{X}^{\top}\bm{\mathcal{Q}} \mathbf{X} + \mathbf{U}^{\top} \bm{\mathcal{R}} \mathbf{U}\right]
	\label{eq:quad-cost}
\end{equation}
where \(\bm{\mathcal{Q}}\) and \(\bm{\mathcal{R}}\) are positive semi-definite weighting matrices. Then, \eqref{eq:quad-cost} can be expanded as
\begin{equation}
	\begin{aligned}
		\mathbb{E} \!  \Big[ \! &\left( \! \bm{\mathcal{A}} \bm{x}_{0} \! + \! \bm{\mathcal{B}}\mathbf{V} \! + \! \left( \! \bm{\mathcal{G}} \! + \! \bm{\mathcal{B}} \! \bm{\mathcal{M}} \! \right) \! \mathbf{W} \! \right)^{\!\top\!} \! \bm{\mathcal{Q}} \! \left(\bm{\mathcal{A}} \bm{x}_{0} \! + \! \bm{\mathcal{B}}\mathbf{V} \! + \! \left( \!  \bm{\mathcal{G}} \! + \! \bm{\mathcal{B}} \! \bm{\mathcal{M}} \! \right) \! \mathbf{W} \! \right) \! \\ 
		&+ \left( \! \bm{\mathcal{M}}\mathbf{W} \! + \! \bm{V}\right)^{\!\top\!} \! \bm{\mathcal{R}} \! \left(\bm{\mathcal{M}}\mathbf{W} \! + \! \bm{V}\right) \! \Big] \\
		= &\left( \bm{\mathcal{A}} \bm{x}_{0} \right)^{\!\top\!} \! \bm{\mathcal{Q}}\bm{\mathcal{A}} \bm{x}_{0} \!+\! 2\left( \bm{\mathcal{A}} \bm{x}_{0} \right)^{\!\top\!} \! \bm{\mathcal{Q}}\left(\bm{\mathcal{G}} + \bm{\mathcal{B}} \bm{\mathcal{M}} \right) \mathbb{E}\left[ \mathbf{W} \right]\\
		&+ 2\left( \bm{\mathcal{A}} \bm{x}_{0} \right)^{\!\top\!} \! \bm{\mathcal{Q}} \bm{\mathcal{B}}\mathbf{V} \!+\! 2 \left( \bm{\mathcal{B}}\mathbf{V} \right)^{\!\top\!}\! \bm{\mathcal{Q}}\left(\bm{\mathcal{G}} + \bm{\mathcal{B}} \bm{\mathcal{M}} \right) \mathbb{E}\left[ \mathbf{W} \right]\\
		&+  \left( \bm{\mathcal{B}}\mathbf{V} \right)^{\!\top\!} \! \bm{\mathcal{Q}}\bm{\mathcal{B}}\mathbf{V} +\mathbb{E}\left[ \mathbf{W}^{\!\top\!} \! \left(\bm{\mathcal{G}} + \bm{\mathcal{B}} \bm{\mathcal{M}} \right)^{\!\top\!} \! \bm{\mathcal{Q}} \left(\bm{\mathcal{G}} + \bm{\mathcal{B}} \bm{\mathcal{M}} \right)\mathbf{W}\right] \\
		&+\mathbb{E}\left[ \mathbf{W}^{\!\top\!} \! \bm{\mathcal{M}}^{\!\top\!}  \bm{\mathcal{R}} \bm{\mathcal{M}}\mathbf{W} \right] + \mathbb{E}\left[\mathbf{W}^{\top}\right] \bm{\mathcal{M}}^{\!\top\!}  \bm{\mathcal{R}} \bm V \\
		&+ \bm V^{\top}\bm{\mathcal{R}}\bm{\mathcal{M}}\mathbb{E}\left[\mathbf{W}\right] + \bm V^{\top} \bm{\mathcal{R}} \bm V.
	\end{aligned}	
\end{equation}

Since we assume standard Gaussian distribution, it holds that \(\mathbb{E}\left[\mathbf{W}\right] = \bm 0\) and \(\mathbb{E}\left[\mathbf{W}^{\top}\mathbf{W}\right] =\bm I\). By applying Property~\autoref{sec:property2}, the expected value of the state and input costs become
\begin{equation}
	\begin{aligned}
		\mathbb{E}\left[\lVert \mathbf{X} \rVert_{\bm{\mathcal{Q}}}^{2} \right]
		&= \left( \bm{\mathcal{A}} \bm{x}_{0} \right)^{\!\top\!} \! \bm{\mathcal{Q}}\bm{\mathcal{A}} \bm{x}_{0} + 2\left( \bm{\mathcal{A}} \bm{x}_{0} \right)^{\!\top\!} \! \bm{\mathcal{Q}} \bm{\mathcal{B}}\mathbf{V}\\
		&+ \! \left( \bm{\mathcal{B}}\mathbf{V} \right)^{\!\top\!} \! \bm{\mathcal{Q}}\bm{\mathcal{B}}\mathbf{V} \!+\! \mathrm{tr}\left(\left(\bm{\mathcal{G}} + \bm{\mathcal{B}} \bm{\mathcal{M}} \right)^{\!\top\!} \! \bm{\mathcal{Q}} \left(\bm{\mathcal{G}} \!+\! \bm{\mathcal{B}} \bm{\mathcal{M}} \right)\right)
	\end{aligned}
\end{equation}
and 
\begin{equation}
	\begin{aligned}
		\mathbb{E}\left[\lVert \mathbf{U} \rVert_{\bm{\mathcal{R}}}^{2}\right] = \bm V^{\top} \bm{\mathcal{R}} \bm V \!+\! \mathrm{tr}\left( \bm{\mathcal{M}}^{\!\top\!}  \bm{\mathcal{R}} \bm{\mathcal{M}} \right).
	\end{aligned}
\end{equation}

\section*{Acknowledgment}
F. M. Barbosa thanks Miriam Zin{\ss}el for the valuable discussions and insights.

\section*{References}
\bibliographystyle{IEEEtran}
\bibliography{reference}

\begin{IEEEbiography}[{\includegraphics[width=1in,height=1.25in,clip,keepaspectratio]{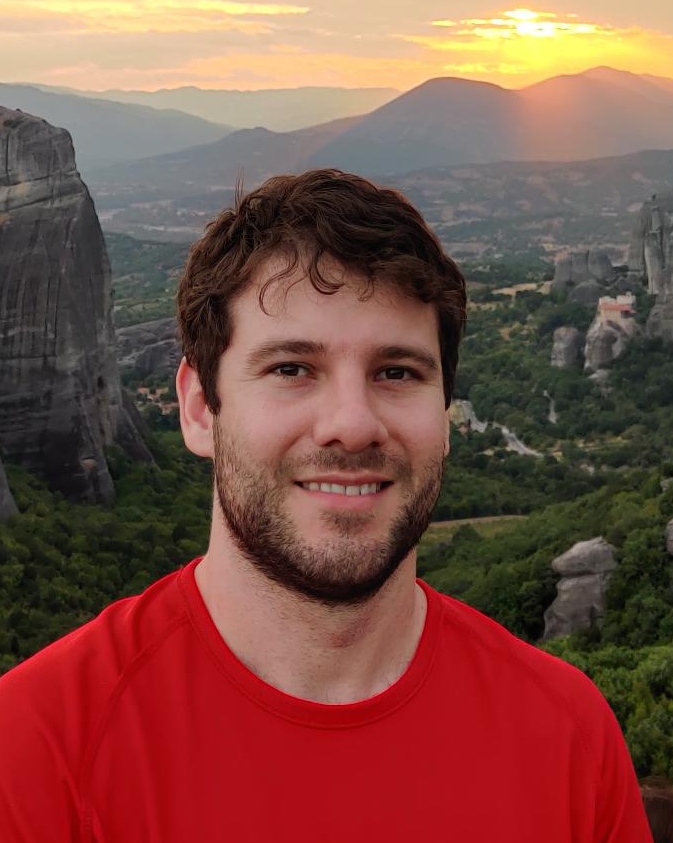}}]{Filipe Marques Barbosa} received the B.Sc. degree in electrical engineering from the Federal University of Uberl{\^a}ndia, Brazil, in 2016, and the M.Sc. degree in dynamical systems from the University of S{\~a}o Paulo, Brazil, in 2018. He is currently working toward the Ph.D. degree in automatic control with the Division of Automatic Control, Department of Electrical Engineering, Link{\"o}ping University, Sweden. 
	
His main research interests lie in optimization for control theory, with particular interest in optimization under uncertainty and its applications in stochastic and robust control.
\end{IEEEbiography}

\begin{IEEEbiography}[{\includegraphics[width=1in,height=1.25in,clip,keepaspectratio]{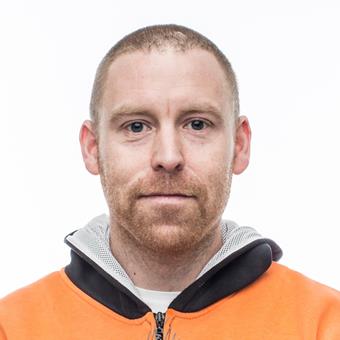}}]{Johan L{\"o}fberg} received the M.Sc. degree in mechanical engineering and the Ph.D. degree in automatic control from Link{\"o}ping University, Link{\"o}ping, Sweden, in 1998 and 2003, respectively. From 2003 to 2006, he was a Post-Doctoral Researcher with ETH Zurich, Z{\"u}rich, Switzerland. He is currently Associate Professor and Docent with the Division of Automatic Control, Department of Electrical Engineering, Link{\"o}ping University. 
	
He is the author of the MATLAB toolbox YALMIP, which is an established tool for optimization for researchers and engineers in many domains. His main research interests include general aspects of optimization in control and systems theory, with a particular interest in model predictive control. Driven by applications in control, he is also more generally interested in optimization under uncertainty and optimization modeling.
\end{IEEEbiography}

\end{document}